\documentclass[12pt]{article}
\pdfoutput=1
\usepackage[utf8]{inputenc}
\usepackage{dsfont}
\usepackage{diagbox}
\usepackage{amsfonts}
\usepackage{mathtools}
\allowdisplaybreaks[4]        
\usepackage{amssymb}
\usepackage{euscript}     
\usepackage{braket}
\usepackage{starfont}
\usepackage{color,soul}         
\usepackage{braket}
\usepackage{tensor}        
\usepackage{amsthm}
\usepackage{graphicx}
\usepackage{slashed}
\usepackage{leftidx}
\usepackage{subfigure}
\usepackage{bbm}
\usepackage{empheq}
\definecolor{outerspace}{rgb}{0.25, 0.29, 0.3}
\definecolor{scarlet}{rgb}{1.0, 0.13, 0.0}
\usepackage[header,title,page,titletoc]{appendix}  
\definecolor{princetonorange}{rgb}{1.0, 0.56, 0.0}
\definecolor{WildStrawberry}{rgb}{1.0, 0.26, 0.64}
\definecolor{rossocorsa}{rgb}{0.83, 0.0, 0.0}
\definecolor{navyblue}{rgb}{0.0, 0.0, 0.5}
\usepackage[numbers,sort&compress]{natbib}  
\usepackage{float}
\usepackage[paper=a4paper,margin=1in]{geometry}
\usepackage{titlesec}

\pdfoutput=1
\usepackage{amsmath}
\usepackage{color}
\usepackage{amsfonts}
\usepackage{amssymb}
\usepackage{graphicx}
\usepackage{geometry}
\usepackage{amssymb,epsfig,subfigure}
\usepackage{amssymb}
\usepackage{hyperref}
\usepackage{comment}
\usepackage[font=footnotesize]{caption}
\usepackage[T1]{fontenc}
\usepackage[utf8]{inputenc}
\usepackage{latexsym}

\usepackage{cancel}

\usepackage[numbers,sort&compress]{natbib}  
\usepackage{float}

\usepackage{dsfont}
\usepackage{diagbox}

\usepackage{mathtools}
\allowdisplaybreaks[4]

\usepackage{euscript}     
\usepackage{braket}
\usepackage{starfont}
\usepackage{color,soul}         
\usepackage{tensor}        
\usepackage{amsthm}

\usepackage{slashed}
\usepackage{leftidx}
\usepackage{bbm}

\makeatletter
\renewcommand\section{\@startsection {section}{1}{\z@}%
                                 {-3.5ex \@plus -1ex \@minus -.2ex}
                                   {2.3ex \@plus.2ex}%
                                   {\normalfont\large\bfseries}}
\renewcommand\subsection{\@startsection{subsection}{2}{\z@}%
                                   {-3.25ex\@plus -1ex \@minus -.2ex}%
                                     {1.5ex \@plus .2ex}%
                                     {\normalfont\bfseries}}
\renewcommand\subsubsection{\@startsection{subsubsection}{3}{\z@}%
                                   {-3.25ex\@plus -1ex \@minus -.2ex}%
                                     {1.5ex \@plus .2ex}%
                                     {\normalfont\itshape}}
\makeatother

\def\pplogo{\vbox{\kern-\headheight\kern -29pt
\halign{##&##\hfil\cr&{\ppnumber}\cr\rule{0pt}{2.5ex}&\ppdate\cr}}}
\makeatletter
\def\ps@firstpage{\ps@empty \def\@oddhead{\hss\pplogo}%
  \let\@evenhead\@oddhead 
}
\def\maketitle{\par
 \begingroup
 \def\thefootnote{\fnsymbol{footnote}}
 \def\@makefnmark{\hbox{$^{\@thefnmark}$\hss}}
 \if@twocolumn
 \twocolumn[\@maketitle]
 \else \newpage
 \global\@topnum\z@ \@maketitle \fi\thispagestyle{firstpage}\@thanks
 \endgroup
 \setcounter{footnote}{0}
 \let\maketitle\relax
 \let\@maketitle\relax
 \gdef\@thanks{}\gdef\@author{}\gdef\@title{}\let\thanks\relax}
\makeatother

\numberwithin{equation}{section}

\newcommand\eea{\end{eqnarray}}
\newcommand\bea{\begin{eqnarray}}

\def\beq{\begin{equation}}
\def\eeq{\end{equation}}

\newcommand{\be}{\begin{equation}}
\newcommand{\ee}{\end{equation}}
\newcommand{\ba}{\begin{align}}
\newcommand{\ea}{\end{align}}
\newcommand{\bg}{\begin{gather}}
\newcommand{\eg}{\end{gather}}
\newcommand{\bseq}{\begin{subequations}}
\newcommand{\eseq}{\end{subequations}}

\usepackage{hyperref}
\hypersetup{
    colorlinks,
    citecolor=rossocorsa,
    filecolor=navyblue,
    linkcolor=navyblue,
    urlcolor=navyblue
}

\begin{document} 

\begin{titlepage}

\begin{center}

\phantom{ }
\vspace{3cm}

{\bf \Large{Generalized Symmetries of the Graviton} }
\vskip 0.5cm
Valentin Benedetti${}^{a}$, Horacio Casini${}^{b}$, Javier M. Mag\'an${}^{c}$
\vskip 0.05in
\small{ ${}^{a}$ ${}^{b}$ \textit{Instituto Balseiro, Centro At\'omico Bariloche}}
\vskip -.4cm
\small{\textit{ 8400-S.C. de Bariloche, R\'io Negro, Argentina}}
\vskip -.10cm
\small{${}^{c}$ \textit{David Rittenhouse Laboratory, University of Pennsylvania}}
\vskip -.4cm
\small{\textit{ 209 S.33rd Street, Philadelphia, PA 19104, USA}}
\vskip -.10cm
\small{${}^{c}$ \textit{Theoretische Natuurkunde, Vrije Universiteit Brussel (VUB) }}
\vskip -.4cm
\small{\textit{ and The International Solvay Institutes}}
\vskip -.4cm
\small{\textit{ Pleinlaan 2, 1050 Brussels, Belgium}}

\begin{abstract}

We find the set of generalized symmetries associated with the free graviton theory in four dimensions. These are generated by gauge invariant topological operators that violate Haag duality in ring-like regions. As expected from general QFT grounds, we find a set of ``electric'' and a dual set of ``magnetic'' topological operators and compute their algebra. To do so, we describe the theory using phase space gauge-invariant electric and magnetic dual variables constructed out of the curvature tensor. Electric and magnetic fields satisfy a set of constraints equivalent to the ones of a stress tensor of a $3d$ CFT. The constraints give place to a group $\mathbb{R}^{20}$ of topological operators that are charged under space-time symmetries. Finally, we discuss 
 similarities and differences between linearized gravity and tensor gauge theories that have been introduced recently in the context of fractonic systems in condensed matter physics.

\end{abstract}
\end{center}

\small{\vspace{3.5 cm}\noindent ${}^{a}\,\,$valentin.benedetti@ib.edu.ar\\
${}^{b}\,\,$casini@cab.cnea.gov.ar\\
${}^{c}\,\,$magan@sas.upenn.edu}

\end{titlepage}

\setcounter{tocdepth}{2}

{\parskip = .4\baselineskip \tableofcontents}
\newpage


\section{Introduction}

Generalized symmetries have been recently studied in many different QFT, string theory,  and condensed matter systems (see for example the nice review talk in the Strings 2021 \cite{string21}). The purpose of this article is to derive and analyze the generalized symmetries of the graviton field in $(3+1)$ dimensions. The graviton field is treated here as a QFT of a free massless helicity two field. 

Before describing our motivations to do so, we start with a brief account of the concept of generalized symmetries from a broad perspective. Generalized symmetries should not be confused with gauge symmetries. The theory of the photon field has an Abelian gauge symmetry, but this symmetry acts trivially in all states and observables of the theory. Still, this gauge symmetry leaves a physical remnant, two generalized symmetries generated by Wilson and 't Hooft lines, under which the 't Hooft and Wilson lines are charged respectively. This notion was first evision in 't Hooft's seminal article \cite{tHooft:1977nqb}, and it was recently put on a broader and firmer concept in \cite{Gaiotto:2014kfa}, where they were called generalized global symmetries.

More recently, in \cite{Casini:2020rgj}, the concept of generalized symmetries has been extended and related to basic properties of local QFT. Given a region $R$ with  algebra $\mathcal{A}(R)$,  generated by local degrees of freedom, the usual commutativity of spatially separated observables (causality) is expressed as
\be \label{caus}
\mathcal{A}(R)\subseteq \mathcal{A}(R')'\;,
\ee
where $R'$ is the causal complement of the region $R$, i.e., the points that are spatially separated from $R$, while $\mathcal{A}'$ is the commutant of $\mathcal{A}$, i.e., the set of operators that commute with $\mathcal{A}$. The proposal of \cite{Casini:2020rgj} is that the non-saturation of~(\ref{caus}) implies the existence of a certain generalized symmetry. See \cite{Review} for a brief review and \cite{Casini:2019kex} for an extensive treatment of the case of global symmetries.  The reason is that for such inclusion to be non-saturated it has to be the case that
\be 
\mathcal{A}_\textrm{max}(R)\equiv\mathcal{A}(R')'=\mathcal{A}(R)\vee \,\{a\}\;,
\ee
for some set $\{a\}$ of operators which are not generated by the local degrees of freedom in $R$. The symbol $\vee$ denotes the generated algebra by two algebras.  Moreover, as described in \cite{Casini:2020rgj}, the non-saturation of~(\ref{caus}) implies a non-saturation of the dual inclusion, namely
\be \label{causp}
\mathcal{A}(R')\subset \mathcal{A}(R)'\;,
\ee
implying that 
\be 
\mathcal{A}_\textrm{max}(R')\equiv\mathcal{A}(R)'=\mathcal{A}(R')\vee \,\{b\}\;,
\ee
for some set $\{b\}$ of dual ``non-locally generated'' operators in the complementary region $R'$. These sets $\{a\}$ and $\{b\}$ are the topological operators effecting a pair of generalized symmetries under which the sets $\{b\}$ and $\{a\}$ are charged, respectively.\footnote{This simple consequence of the approach developed in \cite{Casini:2020rgj} has been argued to be at the root of the absence of generalized symmetries in the bulk of holographic theories \cite{Review}.} Given a theory, the objective is then to find the full set of non-locally generated operators, and compute their algebra and fusion rules. Generalized global symmetries are included in this approach, but more general fusion rules and algebras than those coming from symmetry groups can appear as well. Also, one could envision generalized symmetries charged under space-time symmetries, and we expect this approach to be sensitive to those as well. We will confirm this below.

Given this brief introduction, the objective of this article is to derive and analyze the non-local operators, their fusion,  and their algebra for linearized gravity.

There are several motivations to do so. Let's start with the ones coming purely from QFT. As it is well known, several obstructions and peculiarities appear when trying to construct QFT's with massless particles of spin two. On one hand,  we have the Weinberg-Witten theorem \cite{WEINBERG198059}, which precludes the existence of a stress tensor for these theories. On the other hand, we have the Coleman-Mandula theorem, which precludes the existence of non-trivial mixing of space-time symmetries with internal symmetries.  It would be interesting to deepen on these issues from the point of view of generalized symmetries. In particular, since the gauge constraints of the graviton are related to space-time symmetries, one might envision that the generalized symmetries might be charged under the Poincar\'e group.  We will indeed find the graviton has non-trivial sectors for non-contractible loops, and these sectors are charged under Lorentz symmetries. This, in principle,  may set a counterexample to a putative generalized Coleman-Mandula for these types of symmetries. This example could be accounted for because the theory is free. But it would be giving a further strong argument against potential interacting Lorentz invariant theories of the graviton.
 
From the perspective of QFT alone, we are also motivated to study this problem to better understand the scope of the previous approach to generalized symmetries \cite{Casini:2020rgj}. This approach was analyzed for non-abelian gauge theories in \cite{Pedro}. But we expect the graviton to show various peculiar behaviors that might escape the standard lore of generalized global symmetries \cite{Gaiotto:2014kfa}.

Let's now describe some motivations coming from quantum gravity. Since the appearance of \cite{Polchinski:2003bq} and \cite{Banks:2010zn}, there has been an increasing growth of interest in trying to understand completeness in quantum gravity. Starting from \cite{Banks:2010zn}, and more recently in \cite{Rudelius:2020orz,Casini:2020rgj,Heidenreich:2020tzg,Review}, the issue of completeness has been rightly connected to the absence of generalized symmetries. The previous approach makes this transparent, as reviewed above. Given a QFT with a dual structure of generalized symmetries, to make the theory complete,  we need to introduce a sufficient number of charged operators such that the previous inclusions become saturated, or equivalently, such that the non-locally generated operators become locally generated. Several examples of this completion process are described in \cite{Casini:2020rgj,Review}, where the role of the Dirac quantization conditions is accounted for. In this vein, by studying the generalized symmetries of the graviton, we are also paving the path for a potential understanding of how to complete such a theory and/or what problems or subtleties arise in the process. In this context, we will see that to destroy the non-local operators while keeping within QFT, in principle, we will need to break Poincar\'e invariance dramatically.

Finally, and maybe unexpectedly, there are also motivations coming from condensed matter physics. Recently, some new type of excitations dubbed fractons have been explored, see \cite{OrginalFracton1,OrginalFracton2,OrginalFracton3,OrginalFracton4,Pretko:20161} or the review \cite{Pretko:2020}.  Recent QFT realizations  are described in \cite{Seiberg:2020bhn,Seiberg:2020wsg,Seiberg:2020cxy}, and Ref. \cite{exp1,exp2} describe potential experimental senarios. These are theories without relativistic symmetry. The mobility properties of these fractons are completely determined by the nature of the generalized symmetries associated with the gauge fields to which fractons couple. These generalized symmetries, and their associated topological operators, are typically charged under spacial symmetries, and this is at the root of the restricted mobility properties of fractons. A heuristic connection between the physics of fractons and gravitational physics was laid down in \cite{Pretko:2017}, based on the Hamiltonian and momentum constraints. In this article, by analyzing in detail the generalized symmetries of the graviton, we will be able to make more precise the connections and differences between the physics of gravitons and the physics of the tensor gauge fields that have been associated with fractons.

The article is structured as follows. In section~(\ref{LG}),  we will introduce the theory of the graviton (linearized gravity) and review the derivation of Hamiltonian and momentum constraints, together with their respective charges. These can be written in terms of surface integrals of non-gauge invariant operators. However, they are not responsible for the generalized symmetries. To find generalized symmetries it is necessary to write everything in terms of gauge invariant operators and study the non-saturation of~(\ref{caus}) for such algebra. To do so,  we reformulate the graviton theory in terms of certain electric and magnetic fields, finally re-deriving the formulation of Longo et al \cite{2019}. We use this electromagnetic formulation of the graviton in section~(\ref{TG}) to find the dual sets of generalized charges. These turn out to be one form symmetries charged under the Poincar\'e group (they carry space indices). Using the canonical commutation relations of the graviton we derive the commutation relations of the electric and magnetic graviton fields. With these commutations relations, we find the algebra of topological operators. This is done in a novel way that uses the smearing of the topological operators. In the last section~(\ref{FR}), we discuss the connection between the physics of the graviton and the physics of fractonic systems. We briefly explore the nature of charged operators we need to include to explicitly break the generalized symmetries and complete the graviton theory within the QFT framework. We end with some discussion related to the Coleman-Mandula and the Weinberg-Witten theorems in section~(\ref{DS}).

\section{Linearized gravity \label{LG}}

We will be interested in analyzing the theory of the linearized graviton in flat spacetime. This arises by considering small metric perturbations over a fixed Minkowski background. The metric is then
\be
g_{\mu\nu}=\eta_{\mu\nu}+h_{\mu\nu}\,,\quad ||h_{\mu\nu}||\ll 1\,.
\label{Definition}
\ee
Using this expression one can expand to any desired order any curvature related variable in general relativity. In particular, starting from Einstein's equations in $(3+1)$ dimensions without external sources, the linearized equation of motions are
\be 
 \partial^\lambda \partial_\mu h_{\lambda\nu} +\partial^\lambda \partial_\nu h_{\lambda\mu} - \partial^\lambda \partial_\lambda h_{\mu\nu} - \partial_\mu \partial_\nu h_\sigma^\sigma - \eta_{\mu\nu}(\partial^\lambda \partial_\sigma h_{\lambda\sigma} - \partial^\sigma \partial_\sigma h_\rho^\rho)=0\,.
\label{EquationOfMotion}
\ee
Expanding the curvature to first and second order, we get
\bea
R^{(1)} &=& \partial_\rho \partial_\sigma h^{\rho\sigma} - \partial^\rho \partial_\rho h^{\sigma}_{\,\,\,\sigma} \,,\\
R^{(2)} &=& h^{\lambda \sigma} \left[ \partial_\rho \partial^\rho h_{\lambda\sigma}  -  \partial_\lambda \partial^\rho h_{\rho\sigma}  - \partial_\sigma \partial^\rho h_{\rho\lambda} + \partial_\lambda \partial_\sigma h^{\rho}_{\,\,\,\rho} \right]  +  \frac{3}{4} \partial_\sigma h_{\rho \lambda}\partial^\sigma h^{\rho \lambda} \nonumber\\
&-& \partial^\sigma h_{\sigma \lambda} \partial_\rho h^{\rho\lambda} + \partial^\sigma h_{\sigma \lambda} \partial^\lambda h^{\rho}_{\,\,\,\rho}  -   \frac{1}{2} \partial_\sigma h_{\rho \lambda} \partial^\rho h^{\sigma\lambda} -  \frac{1}{4} \partial^\lambda h^{\sigma}_{\,\,\,\sigma}\partial_\lambda h^{\rho}_{\,\,\,\rho}  \,.
\eea
Therefore,  the previous equations of motion (\ref{EquationOfMotion}) can be recovered from  the quadratic Einstein-Hilbert\footnote{For convenience we will consider $(8 \pi G) =1$ unless otherwise expressed.} action
\be
S_{EH}=\int d^4x \,\sqrt{g} \, R \ \approx \int d^4x \,\left[ \left( 1 + \frac{1}{2} h \right) R^{(1)} + R^{(2)}  \right] \;.
\label{EinsteinHilbert}
\ee
This is the usual Fierz-Pauli action for the graviton \cite{Ortin:2004}, plus some boundary terms that arise from integrating by parts the curvature, namely 
\be 
S_{FP}=\int d^4x \,\left[ -\frac{1}{2}\partial_\mu h_\rho^\rho \partial_\nu h^{\mu\nu}+\frac{1}{2}\partial_\lambda h^{\mu\nu} \partial_\nu h_{\mu\lambda}+\frac{1}{4}\partial_\mu h_\rho^\rho\partial^\mu h_\sigma^\sigma-\frac{1}{4}\partial^\lambda h^{\mu\nu} \partial_\lambda h_{\mu\nu}\right] \,.
\label{FierzPauli}
\ee
\subsection{Gauge symmetry}

The graviton theory has a gauge symmetry. This is not the full diffeomorphism group, as happens for the full non-linear Einstein theory. The reason is that there are diffeomorphisms that destroy the assumed weak perturbation limit defined by~(\ref{Definition}). We need to restrict to the diffeomorphisms that respect that assumption. To linear order,  we can write the corresponding diffeomorphisms as 
\be 
x^\mu \,\,\rightarrow\,\,{ x'}^\mu=x^\mu +\xi^\mu (x)\,,
\ee
where $\xi^\mu (x)$ is of the same order as $h_{\mu\nu}$.
Under this restricted set of diffeomorphims, the perturbation ${h}_{\mu\nu}$ transforms as
\be 
{h'}_{\mu\nu}(x')= {h}_{\mu\nu}(x)+\partial_\mu \xi_\nu(x)+\partial_\nu \xi_\mu(x)\,,
\label{gaugeh}
\ee
which implies that the linearized Christoffel symbols
\be 
\Gamma_{\alpha\beta}^{\mu\,(1)} = \frac{1}{2}\eta^{\mu n} \left(h_{n\alpha,\beta}+ h_{n\beta,\alpha}+ h_{\alpha\beta,n}\right)\,,
\ee
transform as 
\be
{\Gamma'}^{\mu\,\,(1)}_{\alpha\beta} = {\Gamma}^{\mu\,\,(1)}_{\alpha\beta}+\partial_\alpha \partial_\beta \xi^\mu \,.
\ee
On the other hand, the Riemann tensor to first order
\be\label{rie}
R^{(1)}_{\rho\sigma\mu\nu}=\frac{1}{2}\left(\partial_\sigma \partial_\mu h_{\rho\nu} +\partial_\rho \partial_\nu h_{\sigma\mu} - \partial_\rho \partial_\mu h_{\sigma\nu} - \partial_\sigma \partial_\nu h_{\rho\mu}\right)\;,
\ee
is clearly gauge invariant
\be 
{R'}^{\rho\,\,(1)}_{\,\,\, \sigma\mu\nu} = {\Gamma'}_{\nu\sigma,\mu}^{\rho\,\,(1)} -{\Gamma'}_{\mu\sigma,\nu}^{\rho\,\,(1)} =\left( {\Gamma}_{\nu\sigma,\mu}^{\rho\,\,(1)} -{\Gamma}_{\mu\sigma,\nu}^{\rho\,\,(1)}\right)+\left(\partial_\nu \partial_\sigma \partial_\mu \xi^\rho - \partial_\mu \partial_\sigma \partial_\nu \xi^\rho \right)= {R}^{\rho\,\,(1)}_{\,\,\, \sigma\mu\nu} \,.
\ee

We will be interested in the theory of linearized gravitons in flat space, which is a theory of spin two massless particles. The curvature tensor (\ref{rie}), as an operator in the quantum theory, generates the local gauge invariant algebra of the graviton theory.

\subsection{Hamiltonian and momentum constraints and conserved charges}

To derive the non-local operators effecting the generalized symmetries, one could naively follow a similar procedure as in the Maxwell theory. In Maxwell theory,  one can integrate the gauge constraint $\nabla \cdot E =0$ over a volume in space. This gives rise to Gauss law in the absence of charges. Gauss law implies that if we now define a flux operator over an open surface, this operator does not depend on the chosen surface, just on its boundary. This implies that such an operator can be associated with any ring containing the boundary since it commutes with all local gauge invariant operators outside such ring. These surface operators are the 't Hooft loops. For a more detailed description of the relation between these operators and the inclusion~(\ref{caus}), see \cite{Casini:2020rgj}.

Let's quickly go over this method in the graviton theory.  In this case,  the equations of motion of the Lagrange multipliers $h_{00}$ and $h_{0i}$ respectively produce the linearized Hamiltonian and momentum constraints 
\be 
\mathcal{H} = -2G_{00}^{(1)}= \partial_i \partial^i  h^j_{\,\,j}- \partial^i \partial^j h_{ij}   =0\,, \quad  \mathcal{H}_i =-2G_{0i}^{(1)}= -2 \,\partial_j \pi^{ij}   =0\,,
\label{Constraints}
\ee
where $\pi_{ij}$  denotes the canonical momenta associated with the remaining dynamical variables $h_{ij}$. More explicitly we have
\be 
\pi_{ij} =  \frac{1}{2}\left( \dot{h}_{ij} - \partial_i h_{0j} - \partial_j h_{0i} - \delta_{ij} \dot{h}^{k}_{\,\,k} +2\, \delta_{ij} \partial^k h_{0k}  \right)\,.
\label{Momenta}
\ee
For completeness, in appendix~\ref{ADM} we include a derivation of the constraints from the linearized ADM theory, though this will not be required in what follows. 

At the quantum level, the canonical dual variables $h_{ij}$  and $\pi_{ij}$ satisfy the equal time commutation relations
\be 
\left[h_{ij}(x), \pi^{kn}(y)\right]= \frac{i}{2}\left(\delta^k_i \delta^n_j  + \delta^n_i \delta^k_j \right)\delta(x-y)\;.
\label{Canonical}
\ee
From these commutation relations, it is straightforward to find that the smeared constraints produce the action of linearized diffemorphisms (\ref{gaugeh}) over the canonical variables
\bea
&{}&\left[h_{ij}(x) , \int d^3y\, \mathcal{H}^k (y)\,\xi_k (y)\, \right] = i\left( \partial_i \xi_j(x) +  \partial_j \xi_i (x)\right)\,,\label{TransformField}\\
&{}&  \left[\pi_{ij}(x) , \int d^3y\, \mathcal{H} (y)\,\xi_0 (y)\, \right] = i\left(\partial_i \partial_j -\delta_{ij} \partial^2 \right)\xi_0(x)\;. \label{TransformMomenta}
\eea
As with the Maxwell field, one might be interested in finding the charges associated with these constraints, i.e with diffeomorphism invariance.  Considering Noether's second theorem for linearized diffeomorphisms\footnote{This requires the addition of a Gibbons-Hawking-York boundary term to secure a well-defined variational problem with Dirichlet boundary conditions.} it is possible to derive the associated charges more rigorously  \cite{Barnich:2001,Avery:2015}. These charges can be written as integrals of the constraints over a region $V $ defined in a Cauchy slice, or as fluxes over the closed boundary $\partial V$ of such region. More explicitly:
\newpage
 \begin{align}  
P^0&=-\frac{1}{2}\int_Vd^3x \,\mathcal{H}=\frac{1}{2}\int_{\partial V}  dS_n \left( \partial_j h^{nj}  - \partial^n h^j_{\,\,j}  \right)\,,\label{P0}\\
P^i&=-\frac{1}{2}\int_V  d^3x \, \mathcal{H}^i =\frac{1}{2} \int_{\partial V}  dS_n \left[  \partial^n h_{0}^{\,\,i}- \dot{h}^{ni} - \delta^{ni}(  \partial_j h_{0}^{\,\, j}- \dot{h}^j_{\,\,j} )  \right]\,,\nonumber\\
J^{0i}& =-\frac{1}{2}\int_V  d^3x \, (x^i \mathcal{H} - x^0\mathcal{H}^i) \nonumber \\ & =\frac{1}{2}\int_{\partial V}  dS_n \left[x^i ( \partial^j h_{jn} - \partial^n h^j_{\,\,j})  -  h^{ni} + \delta^{ni} h^j_{\,\,j} - x^0(\partial^n h_{0}^{\,\,i} -\dot{h}^{ni}-\delta^{ni}\partial^j {h}^0_{\,\,j}+\delta^{ni}\dot{h}^j_{\,\,j} ) \right]\,,  \nonumber \\ 
J^{ij} & =-\frac{1}{2}\int_V d^3x \, (x^j \mathcal{H}^i- x^i \mathcal{H}^j )\nonumber \\ &=\frac{1}{2}\int_{\partial V} dS_n \left[x^j( \partial^n h_{0}^{\,\,i}-\dot{h}^{ni}-\delta^{ni}\partial_m{h}^0_{\,\,m}+\delta^{ni}\dot{h}^m_{\,\,m} )-    \delta^{jn} h_0^{\,\,i}-  (i\leftrightarrow j ) \right]\,. \nonumber
\end{align}
These are the well-known ``Poincar\'e charges''. These are precisely the ones originally proposed in \cite{Weinberg:1972kfs} via the construction of the effective graviton energy-momentum tensor (when one considers the full general relativity equations to be valid).  These Poincar\'e charges can be seen to satisfy the Poincar\'e algebra \cite{Abbott:1981},  which can be thought of as a consequence of linearization of the well-known Dirac-Schwinger algebra of the naked non-linear ADM constraints  \cite{DeWitt:1967}.

But here,  we see the first deviation from the Maxwell theory. While in Maxwell theory the charges are written as fluxes of gauge invariant local operators (electric and magnetic fields), in gravity these Poincar\'e conserved charges that arise by integrating the Hamiltonian and Momentum constraints are fluxes of gauge non-invariant local operators. While the constraints and associated charges themselves are gauge invariant, this feature prevents us from extracting from them Wilson and 't Hooft loops in the way one does for the Maxwell field. More concretely, if we integrate the previous fluxes over open surfaces we get line operators which are not gauge invariant, and therefore they do not generate symmetries of the physical gauge invariant Hilbert space in which the constraints are fixed to zero.

Equivalently, the difference between the Maxwell and the present gravity case shows up in that the effective graviton energy-momentum tensor does not present diffeomorphism invariance (except on the boundary of the manifold where gauge transformations vanish).  Due to the Weinberg-Witten theorem \cite{WEINBERG198059},  there is no gauge invariant, conserved and Lorentz covariant energy-momentum tensor that produces these charges in the spin 2 theory. However, note that linearized gravity theory exhibits conformal symmetry at least on-shell over a Minkowski background as proven in \cite{2019}. Some discussions about this fact can be found in \cite{Conformal1,Conformal2} or in the context of entanglement entropy calculations in \cite{Valentin,David:2021}.

\subsection{Gauge invariant phase space variables of linearized gravity}

The formulation of the linearized graviton field in terms of phase space gauge invariant variables makes the problem of finding the generalized symmetries easier to understand. One seeks to find gravitational electric and magnetic fields, playing analog roles like the ones of the Maxwell field. In the case of the Maxwell field, the full algebra of the theory is generated by such fields. So let's start by looking for a set of gauge invariant operators generating the graviton algebra.

In contrast to what is expected in full quantum gravity, the theory of the graviton field in Minkowski space contains gauge invariant local operators.  As mentioned above, these can be constructed from the linearized Riemann tensor\footnote{From now on we will suppress the order index and write $R_{\mu\nu\rho\sigma}^{(1)}$ as  $R_{\mu\nu\rho\sigma}$.}, defined in~(\ref{rie}). This tensor generates the full local algebra. Also, it exhibits the same symmetries of the non-linear Riemann tensor. One path to writing the theory in terms of electric and magnetic fields uses the fact that the Riemann tensor can be understood as the curvature of a local Lorentz gauge connection, the spin connection.\footnote{For a complete account of this approach and its history see \cite{Ortin:2004}.} This suggests defining the following electric and magnetic fields
\be 
\tilde{E}_i^{\,\,\alpha\beta}=-R_{0i}^{\quad \alpha\beta}\,,\quad \tilde{B}_i^{\,\,\alpha\beta}=\frac{1}{2}\varepsilon_{ijk} R_{jk}^{\quad \alpha\beta}\,.
\label{ElectricMagneticNatural}
\ee
However, as it is well known, the gauge theory formulation of General Relativity differs in many respects from Maxwell. For the present concerns, and in contrast to Maxwell, using the Heisenberg equation for the canonical momenta $\pi_{ij}$ (or the equations of motion in the classical picture), one observes that these fields are not at all independent. Indeed
\be 
\tilde{E}_i^{\,\,0a} =  \varepsilon_{ijk} \, \delta^{jn} \tilde{B}_n^{\,\,ka} \,,\quad \tilde{B}_i^{\,\,0a} = - \varepsilon_{ijk} \, \delta^{jn }  \tilde{E}_n^{\,\,ka}\,.
\label{ElectricMagneticNaturalRelations}
\ee
In consequence,  the on-shell degrees of freedom of the linearized Riemann tensor can be better described by the choice of electric and magnetic operators presented in \cite{2019}. These are
\be 
E_{ij} =- R_{0i0j}\,,\quad B_{ij} = \frac{1}{2}\varepsilon_{iab} R^{\,ab}_{\,\,\,\,\, 0j}\;.
\label{ElectricMagnetic}
\ee
They provide an independent set of local gauge invariant fields generating the graviton algebra. Importantly, they inherit from the Riemann tensor the following  symmetry properties
\be 
E_{ij}=E_{ji}\,, \quad B_{ij}=B_{ji}\,,\quad E^{i}_{\,\,\,i}=0\,,\quad B^{i}_{\,\,\,i}=0\,,
 \label{Simmetries}
 \ee
 and obey a generalized version of Maxwell equations
 \be
\partial^j E_{ij} =0 \,,\quad \partial^j B_{ij} =0\,,
\label{HigherMaxwellDivergence}
\ee
\be
 \varepsilon_{ink} \partial^n E^{k}_{\,\,\,j} = - \dot{B}_{ij}\,, \quad  \varepsilon_{ink} \partial^n B^{k}_{\,\,\,j} =  \dot{E}_{ij}\;.
\label{HigherMaxwellRotor}
\ee
Using expression~(\ref{rie}), these  electric and magnetic fields can be described in terms of the canonical variables as
 \be 
 E_{ij}= \frac{1}{2} \left( \partial^k \partial_k h_{ij}  +\partial_i \partial_j h^{k}_{\,\,\,k}   - \partial_i \partial^k h_{kj} -  \partial_j \partial^k h_{ki} \right)
  \,,\quad
 B_{ij}= \varepsilon_{ink}\,\partial^k \left(\pi^{n}_{\,\,\,j} -\delta^n_j \frac{\pi}{2} \right)\,,
 \label{ElectricMagneticCanonical}
 \ee
 with $\pi=\pi_j^j$. 
These expressions allow us to interpret the divergence-free and traceless properties of the electric fields as a consequence of the Hamiltonian constraint
 \be 
 \partial^j E_{ij} = \frac{1}{4} \partial_i\mathcal{H} \,, \quad E^{i}_{\,\,\,i}= \frac{1}{4}\mathcal{H} \,, \quad \varepsilon_{ijk}E^{ij}=0\;,
 \label{ce}
 \ee
while the divergence-free and  symmetry properties of the magnetic field come from the momentum constraint
 \be 
 \partial^j B_{ij} =-\frac{1}{2}\varepsilon_{ink} \partial^k\mathcal{H}_n \,, \quad B^{i}_{\,\,\,i}=0\,, \quad \varepsilon_{ijk}B^{ij}=\frac{1}{2}\mathcal{H}_k =0\,.
 \label{cb}
\ee 
Although the graviton theory and Maxwell theory look quite similar in this formulation, there is a key difference when quantizing the theory. The commutation relations between electric and magnetic fields turn out to be very different. In the present case, using the canonical commutation relations (\ref{Canonical}), and the convenient expressions (\ref{ElectricMagneticCanonical}), we obtain\footnote{The commutation relations must be gauge invariant.  A simpler exercise is to compute these commutations relations using Gupta's quantization scheme \cite{Gupta:1952,Bracci:1972} in harmonic gauge. The result is the same.} 
\begin{align}
\left[E_{ij}(x),B_{kl}(y)\right] & = \frac{i}{4}\varepsilon_{kab} \left[\delta_{ia} (\partial_j  \partial_l -\delta_{jl}\partial^2) \right. \label{ElectricMagneticCommutator}\\  &\quad + \delta_{ja} \left.(\partial_i \partial_l-\delta_{il}\partial^2) -  \delta_{la} (\partial_i \partial_j - \delta_{ij} \partial^2)   \right]\,\partial_b\,  \delta(x-y)\,.\nonumber 
\end{align}
These expressions obey all the constraints (\ref{Simmetries}) and (\ref{HigherMaxwellDivergence}). Indeed they can be rewritten in a more symmetric manner
\begin{align}
 \left[E_{ij}(x),B_{kl}(y)\right] =&   \frac{i}{8} \left[\varepsilon_{kib}(\partial_j  \partial_l -\delta_{jl}\partial^2)+  \varepsilon_{kjb} (\partial_i \partial_l-\delta_{il}\partial^2)   \right.  \label{ElectricMagneticCommutatorSymmetric}\\
 & \quad+ \left.  \varepsilon_{lib} (\partial_j  \partial_k -\delta_{jk}\partial^2) +  \varepsilon_{ljb}(\partial_i \partial_k-\delta_{ik}\partial^2)\right]\,\partial_b\,  \delta(x-y) \;.  \nonumber
\end{align}
In this Maxwellian formulation, the local algebras are generated by electric and magnetic fields $E_{ij}, B_{ij}$, obeying the equations of motion (\ref{HigherMaxwellDivergence}) and (\ref{HigherMaxwellRotor}), constraints (\ref{Simmetries}), and commutators (\ref{ElectricMagneticCommutatorSymmetric}). The analysis of generalized symmetries is transparent in this formulation as we show in section (\ref{TG}).

\subsection{Duality for the linearized gravitational field \label{DualSection}}
Further study of the generalized symmetries of linearized gravity will require us to understand duality transformation involving a symmetric second rank tensor $h_{\mu\nu}$. The corresponding duality,  analogous to the electromagnetic one, was described in detail in \cite{Casini:2003}. Here we present a review of the useful aspects. We start by writing the action in terms of a ``parent'' Lagrangian
\be
S=\int d^4 x \, \left[ \frac{1}{16}T_{(\alpha\beta)\mu}T^{(\alpha\beta)\mu}+  \frac{1}{8}T_{(\alpha\beta)\mu}T^{(\alpha\mu)\beta}+ \frac{1}{4}T_{(\alpha\beta)\mu}\varepsilon^{\alpha\beta\nu\rho}\partial_\nu h^{\mu}_{\,\,\,\, \rho}\right]\,, \label{Dualaction}
\ee
where the field $T_{(\alpha\beta)\mu}$ is an anti-symetric tensor in the indices  $\alpha\leftrightarrow\beta $ and also has zero trace $T_{(\alpha\beta)}^{\quad\alpha}=0$. We can solve the equation of motion of  $T_{(\alpha\beta)\mu}$ in terms of $h_{\mu\nu}$ so as to obtain 
\be  
T_{(\alpha\beta)\mu} =\varepsilon_{\alpha\beta\mu\lambda}(\partial_\sigma h^{\sigma\lambda}-\partial^\lambda h^\sigma_{\,\,\, \sigma}) - \varepsilon_{\alpha\beta}^{\quad\sigma\nu}\partial_\sigma h_{\nu \mu} \label{DualTh}\,.
\ee
Note that by replacing (\ref{DualTh}) in (\ref{Dualaction}) we can recover the expected Fierz-Pauli action (\ref{FierzPauli}) for the original field. Additionally, in (\ref{Dualaction})  the field $h_{\mu\nu}$ appears as a Lagrange multiplier yielding, in the absence of sources, the constraint 
\be
\partial_\sigma \left(\varepsilon^{\alpha\beta\sigma\mu}T_{\alpha\beta}^{\quad\nu} + \varepsilon^{\alpha\beta\sigma\nu}T_{\alpha\beta}^{\quad\mu} \right)=0\,.
\ee
The corresponding solution is given by a pair of dual fields $\tilde{h}_{\mu\nu}$ and $\omega_{\mu\nu}$ as
\be 
T_{(\alpha\beta)\mu}= (\partial_\alpha \omega_{\mu\beta}-\partial_\beta \omega_{\mu\alpha}) + (\partial_\alpha \tilde{h}_{\mu\beta}-\partial_\beta \tilde{h}_{\mu\alpha})\,,\quad \omega_{\mu\nu}=-\omega_{\nu\mu}\,,\quad  \tilde{h}_{\mu\nu}= \tilde{h}_{\nu\mu} \label{DualThw}\,.
\ee
The action (\ref{Dualaction}) writes in terms of the symmetric and anti-symmetric fields as
\begin{align}
S=\int d^4x & \left[ \frac{1}{4}\partial^\alpha \tilde{h}_{\mu\nu} \partial_\alpha \tilde{h}^{\mu\nu} -  \frac{1}{3}\partial^\mu \tilde{h}^{\mu\nu} \partial^\alpha \tilde{h}_{\alpha\nu}-\frac{1}{12}\partial^\mu \tilde{h}^{\nu}_{\,\,\nu} \partial_\mu\tilde{h}^{\alpha}_{\,\,\alpha}\right.\nonumber \\ & \left.  +\frac{1}{6}\partial_\mu \tilde{h}^{\alpha}_{\,\,\alpha} \partial_\nu  \tilde{h}^{\mu\nu} -  \frac{1}{3}\partial^\mu \tilde{h}^{\mu\nu} \partial^\alpha \omega_{\nu\alpha}+\frac{1}{6}\partial^\mu \omega^{\nu\mu} \partial^\alpha \omega_{\nu\alpha}\right] \,.
\end{align}
This has a dual gauge symmetry that can be written as 
\begin{align}
{\omega'}^{\mu\nu}(x')= & {\omega}^{\mu\nu}(x)-\partial^\mu \tilde{\xi}^\nu(x)+\partial^\nu  \tilde{\xi}^\mu(x)\,,\\
\tilde{h}'{}^{\mu\nu}(x')= & \tilde{h}{}^{\mu\nu}(x)+\partial^\mu  \tilde{\xi}^\nu(x)+\partial^\nu  \tilde{\xi}^\mu(x)\,.
\end{align}
Useful identities can be obtained from the duality relation among the potential fields  which comes from (\ref{DualTh}) and (\ref{DualThw}).  For instance, we may write
\be 
\varepsilon_{\alpha\beta\mu\lambda}(\partial_\sigma h^{\sigma\lambda}-\partial^\lambda h^\sigma_{\,\,\, \sigma}) - \varepsilon_{\alpha\beta}^{\quad\sigma\nu}\partial_\sigma h_{\nu \mu}=(\partial_\alpha \omega_{\mu\beta}-\partial_\beta \omega_{\mu\alpha}) + (\partial_\alpha \tilde{h}_{\mu\beta}-\partial_\beta \tilde{h}_{\mu\alpha})\,.
\label{DualPotentials}
\ee
This can be shown to give a duality transformation over the Riemann tensor 
\be 
\tilde{R}_{\mu\nu\rho\sigma}=\frac{1}{2}\varepsilon_{\mu\nu\alpha\beta}R^{\alpha\beta}_{\,\,\,\rho\sigma}\,, 
\ee
which interchanges the electric and magnetic fields as 
\be 
E_{ij}=\frac{1}{2}\varepsilon_{iab}\,\tilde{R}^{ab}_{\,\,\,0j}\,,\quad B_{ij}=-\tilde{R}_{0i 0j}\,.
\ee
In addition, a particularly interesting consequence of (\ref{DualPotentials}) is that we can write the magnetic field as a double rotor
\be 
B_{ij}=\frac{1}{2}\varepsilon_{iab}\varepsilon_{jcd} \partial^b \partial^d \tilde{h}^{ac}\,,
\label{DualityB}
\ee
where we have used the canonical definition (\ref{ElectricMagneticCanonical}) and  the expression (\ref{DualPotentials}) contracted with a spatial Levi-civita tensor as
\be
\frac{1}{2}\varepsilon^{abj}T_{(ab)i}= \dot{h}_{i}^{\,\,j} -\partial^j h_{0i}- \delta_{i}^{j}\left(\dot{h}_{aa} -\partial_j h_{aa}\right)= \varepsilon^{abj}\partial_b \left(\tilde{h}_{ia}+\omega_{ia}\right)\,.
\ee
Also, as one would expect from duality, we can use (\ref{ElectricMagneticCanonical}) to obtain
\be 
E_{ij}=-\frac{1}{2}\varepsilon_{iab}\varepsilon_{jcd} \partial^b \partial^d h^{ac}\;.
\label{DualityE}
\ee
when the Hamiltonian constraint\footnote{A factor of $\delta_{ij}\mathcal{H}=0$ should be added to arrive at this expression} and the equations of motion are satisfied.

\section{Topological operators for the graviton}\label{TG}

Having found a formulation of linear gravity in terms of a set of local gauge invariant electric and magnetic fields, we can now proceed to find operators that violate duality in a certain region. From the generalized Maxwell equations~(\ref{HigherMaxwellDivergence}) and~(\ref{HigherMaxwellRotor}), we expect these operators to violate duality on a ring, as in conventional gauge theories \cite{Casini:2020rgj}. Below we confirm this is the case. But we might also naively expect that these operators will only be given by the fluxes of the electric and magnetic fields, as in the  Maxwell case. In other words, we might expect that the fluxes come by looking at the divergence-less field equations as constraints, leading to the usual conserved charges. Although this is partially correct, it misses several charges.   The difference is that for the graviton theory we need to add also the constraints that descend from the metric formulation, which are
\be 
E_{ij}=E_{ji}\,, \quad B_{ij}=B_{ji}\,,\quad E^{i}_{\,\,\,i}=0\,,\quad B^{i}_{\,\,\,i}=0\,.
 \ee
We are going to see that by taking into account these constraints, we will find an enlarged set of conserved charges. Also, in the last section, we will see that adding these constraints suggests that, in a more precise sense, the correct analogy is not between the graviton and the Maxwell field but between the graviton theory and fractonic systems.

\subsection{A new set of conserved charges}

The traceless and symmetry properties of electric and magnetic fields, together with the divergence-less field equations, suggest to better think of these fields as ``stress tensors'' $T_{ij}$ of euclidean conformal field theories in three dimensions. Such a stress tensor satisfies exactly
\be 
T_{ij}=T_{ji}\,, \quad  T^{i}_{\,\,\,i}=0\,, \quad \partial^i T_{ij}=0\;.
\ee
And we know that given such stress tensor, the conserved charges are in one-to-one correspondence with the generators of the conformal group. We just need to contract the stress tensor with the vector fields generating the associated conformal isometries. Using this intuition one can indeed verify that we have an enlarged set of divergence free gauge invariant\footnote{For a linearized diffeomorphism we have that $x^\mu\to x^\mu + \xi^\mu$. But since the electric and magnetic fields are already first order in the perturbation, the non-invariance of these charges comes at second order, and it can be neglected in the linearized theory.} vector field operators. These are
 \begin{align}
 & B^P_i = B_{ji}a^j \,,\quad B^J_i = -B_{ij}s^{j n}x_n\,,\quad B^D_i  =\kappa B_{ij}x^j \,,\quad B^K_i  = B_{ij}(b^j x^2 - 2 x^j  b\cdot x)\,,
\label{MagneticCFT}\\
& E^P_i = E_{ji}\tilde{a}^j \,,\quad E^J_i =- E_{ij}\tilde{s}^{j n}x_n\,,\quad E^D_i  = \tilde{\kappa} E_{ij}x^j \,,\quad E^K_i = E_{ij}(\tilde{b}^j x^2 - 2 x^j  \tilde{b}\cdot x)\,. \label{ElectricCFT}
 \end{align}
We have that $\kappa$, $a_i$,  $b_i$ and $s_{ij}$ are arbitrary scalar, vectors and anti-symmetric tensor respectively. The same applies to $\tilde{\kappa}$, $\tilde{a}_i$,  $\tilde{b}_i$ and $\tilde{s}_{ij}$. The labels $P,J,D,K$ stand in analogy to  translations, rotations, dilatations and special conformal transformations respectively. Indeed, using the generalized Maxwell equations, together with the constraints we have that
\newpage
\bea
\partial^i E^P_i  &=&  (\partial^i E_{ij})\,\tilde{a}^j  = 0\,, \\
\partial^i E^J_i &=& E_{ij} \tilde{s}^{ij} -(\partial^i E_{ij}) \tilde{s}^{j n}x_n   =0\,, \quad  \nonumber \\
\partial^i E^D_i &=& \tilde{\kappa} (\partial^i E_{ij}) x^j  + \tilde{\kappa}  E^i_{\,\,\,i} =0 \,, \nonumber \\
\partial^i E^K_i  &=& (\partial^i E_{ij})(\tilde{b}^j x^2 - 2 x^j  \tilde{b}\cdot x) + 2  E_{ij} (\tilde{b}^j x^i - \tilde{b}^i  x^j) -  2 (\tilde{b}\cdot x)E^i_{\,\,\,i} =0\,.  \nonumber
\eea

\begin{figure}[t]
\includegraphics[width=.4\linewidth]{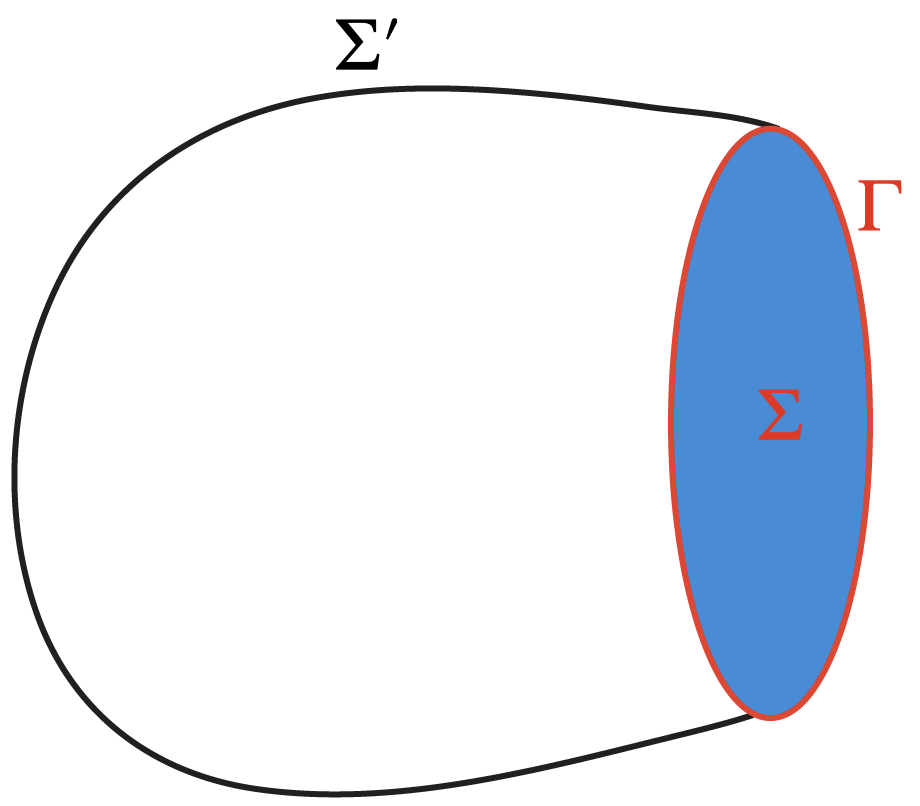}
\centering
\caption{Possible choices $\Sigma$ and $\Sigma'$ for fixed $\Gamma$.}
\label{CurveSigma}
\end{figure}

The previous enlarged divergence-free equations and their magnetic counterparts tell us that that the theory has the following set of conserved fluxes\footnote{When we consider the dimensionless $h_{\mu\nu}$ the Poincare Charges (\ref{P0}) are actually suppressed by a factor $(8\pi G)^{-1}$. This is not the case for the electric and magnetic fluxes (\ref{MagneticFluxes}) and (\ref{ElectricFluxes}).}
\bea
\Phi^B_P=\int_{\Sigma} B^P_i   dS_i \,,\quad \Phi^B_J=\int_{\Sigma} B^J_i   dS_i \,,\quad \Phi^B_D=\int_{\Sigma} B^D_i   dS_i \,,\quad \Phi^B_K=\int_{\Sigma} B^K_i   dS_i\,, \label{MagneticFluxes}\\
\Phi^E_P=\int_{\Sigma} E^P_i   dS_i \,,\quad \Phi^E_J=\int_{\Sigma} E^J_i   dS_i \,,\quad \Phi^E_D=\int_{\Sigma} E^D_i   dS_i \,,\quad \Phi^E_K=\int_{\Sigma} E^K_i   dS_i\,,  \label{ElectricFluxes}
\eea
where the surface of integration is any two-dimensional open surface bounded by a certain ring-like closed boundary.  These flux operators commute with all local operators outside the ring. The reason is the same as in Maxwell theory, we can just move the surface $\Sigma$ of the definition of the topological operator to another one $\Sigma'$ with the same boundary $\partial \Sigma=\partial \Sigma'=\Gamma$ and the operator is unchanged, see figure \ref{CurveSigma}.  Therefore, local operators lying on $\Sigma$ commute with the fluxes by causality,  because we can move the flux away from the support of the operator.

Crucially, as explained in \cite{Casini:2020rgj}, even if they can be associated with ring-like regions by this argument, they are not locally generated in the ring.  In particular, they fail to commute with some of the operators that are non-local in the complementary ring.  We now discuss these issues in detail.

We want the fluxes of the conserved vectors (\ref{MagneticCFT}) and (\ref{ElectricCFT}) to be dimensionless charges. Taking into account that $B_{ij}, E_{ij}$ have dimension three in energy we have $a^i, \tilde{a}^i$ have dimensions of length, $s_{ij}, \tilde{s}_{ij}, \kappa, \tilde{\kappa}$ are dimensionless, and $b_i, \tilde{b}_i$ have dimensions of energy.

\subsection{Algebras of topological operators for the Maxwell field}

Given the set of conserved fluxes and their associated topological operators, the aim now is to compute their algebra. To warm up, and because we are going to introduce new techniques (to our knowledge) adapted to smeared non-local operators to compute commutations relations between non-local operators, we will start by recovering the well-known results presented in \cite{tHooft:1977nqb,Polyakov:1988,Witten:1988}  for the Maxwell field.

The topological operators in the Maxwell theory are known to be Wilson and 't Hooft loops, $W$ and $T$. These are gauge invariant operators that can be obtained by exponentiating the magnetic and electric flux respectively. They are typically defined as singular infinitely thin loops. In this abelian theory, we can define a smeared version of them in the following way.\footnote{Defining smeared versions of loop operators in non-abelian gauge theories turns out to be a particularly difficult problem, see \cite{Pedro} for a recent account from the present perspective.} We start with
\be 
W=e^{i\Phi^B}=e^{i\int d^4 x \,A_\mu J^\mu}\,,\quad \partial_\mu J^\mu =0 \,,
\label{WilsonLoop}
\ee
where $J$ is a conserved current of compact support. These two imposed conditions on the current (smearing function) ensure gauge invariance of the previous operator
\be 
A_\mu \to A_\mu + \partial_\mu \Lambda  \quad \Rightarrow \quad W\to We^{i\int d^4 x \,(\partial_\mu \Lambda) J^\mu} = We^{-i\int d^4 x \,\Lambda (\partial_\mu  J^\mu)} = W\,.
\ee
Now we further assume the support of $J$ is restricted to a region of space-time $R$ which is the domain of dependence of a spatial region containing a non-contractible circle. To be concrete,  we assume the topology of $R$ is $S^1\times \mathbb{R}^3$.  Because of the current conservation, the flux over a 3-dimensional surface $\Sigma$ that cuts the ring $R$ once is independent of the particular $\Sigma$. This flux defines a charge
\be 
q=\int_\Sigma dS_\mu \, J^\mu =\int_\Sigma d\sigma \,n_\mu\, J^\mu\,.
\label{ChargeWilson}
\ee
The vector $n^\mu$ is the normal to $\Sigma$. The claim is that $W$ is a smeared Wilson loop with dimensionless charge $q$. We will confirm this by direct computation in a moment.

An equivalent rewriting of this operator is in terms of the magnetic field. In fact
\be 
\Phi^B=\int d^4 x \,A_\mu J^\mu = \frac{1}{2}\int d^4x\, \omega_{\mu\nu}F^{\mu\nu} \,,
\label{MagneticFluxW}
\ee
where $\omega_{\mu\nu}$ is any 2-form that obeys 
\be
    J_\mu = \partial^\nu \omega_{\mu\nu} \,.
 \ee

Analogously, the smeared t’Hooft loop $T$ for the Maxwell field can be defined from the flux of the electric field
\be 
T= e^{i\Phi^E} = e^{\frac{i}{2}\int d^4x\, \tilde{\omega}_{\mu\nu}(\star F^{\mu\nu})}\,.
\label{ThooftLoop}
\ee
It can also be written in a dual way to the Wilson loop, using the dual gauge field, namely
\be
\Phi^E = \frac{1}{2}\int d^4x\, \tilde{\omega}_{\mu\nu}(\star F^{\mu\nu})=\int d^4x\, \tilde{A}_\mu \tilde{J}^\mu\;,
\ee
where we have that $\star F=d\tilde{A}$ and $ \tilde{J}_\mu = \partial^\nu \tilde{\omega}_{\mu\nu}$. If the current $\tilde{J}$ has support on a ring, the monopole charge of the smeared t’Hooft loop can be measured by integrating the flux of this dual current over a 3-dimensional surface $\tilde{\Sigma}$ with normal vector $\tilde{n}_\mu$ as
\be 
g=\int_{\tilde{\Sigma}} dS_\mu \, \tilde{J}^\mu =\int_{\tilde{\Sigma}} d\sigma\, \tilde{n}_\mu\, \tilde{J}^\mu\,.
\label{ChargeThooft}
\ee
\begin{figure}[t]
\includegraphics[width=.5\linewidth]{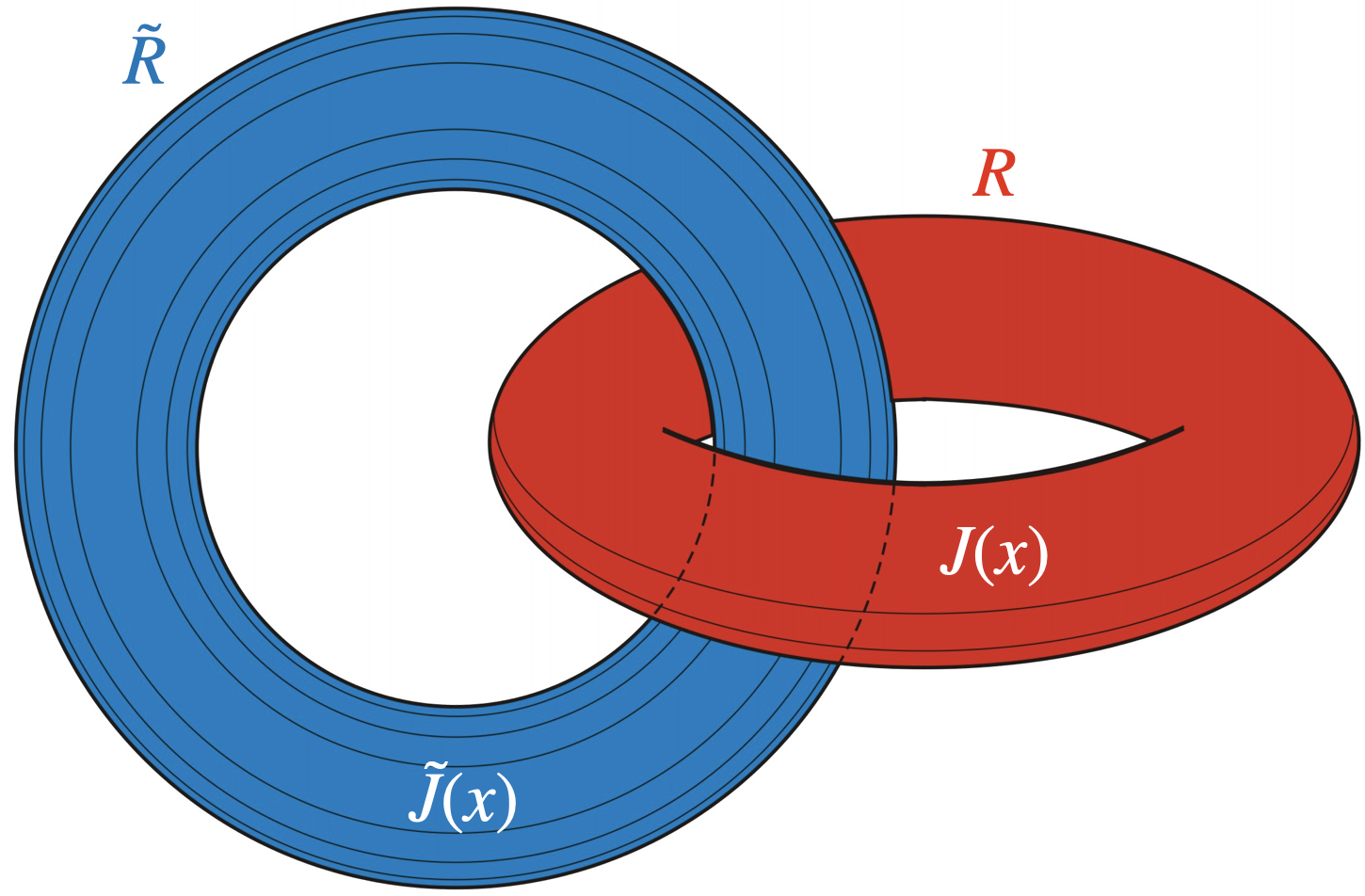}
\centering
\caption{ The ring-like regions $R$ and $\tilde{R}$ are the support of the currents $J(x)$ and $\tilde{J}(x)$.}
\label{rings}
\end{figure}

We are now ready to compute the commutator $\left[\Phi_B, \Phi_E\right]$. The interesting case is when the dual currents $J$ and $\tilde{J}$ have support on linked rings as shown in figure \ref{rings}.
 First, we choose the smearing functions  to have spatial compact support over the rings of interest and also to obey 
\be 
J_i(t,x)\propto \delta(t-t_0) \,,\quad \tilde{J}_i(t,x)\propto \delta(t-t_0) \,,\quad J_0(t,x)= 0 \,,\quad\tilde{J}_0(t,x)= 0\,.
\label{Condition1}
\ee
This implies that the 2-form $\omega_{\mu\nu}$ has zero divergence and spatial components proportional to the delta function 
\be 
\omega_{ij}(t,x)\propto \delta(t-t_0)\,, \quad \tilde{\omega}_{ij}(t,x)\propto \delta(t-t_0)\,,\quad \partial^\nu \omega_{0\nu}(t,x)=0\,, \quad \partial^\nu \tilde{\omega}_{0\nu}(t,x) =0 \,.
\label{Condition2}
\ee
However, we can use the remaining freedom available to pick the stronger condition
\be 
\omega_{0\mu}(t,x)=0\,,\quad\tilde{\omega}_{0\mu}(t,x)=0\,.
\label{Condition3}
\ee
We now allow ourselves a slight abuse in the notation and write $ \omega_{ij}\to\omega_{ij}\, \delta(t-t_0)$ and $(t_0,x)\to x$. We also note that $F_{ij}=\epsilon_{ijk}B^k$ and $\star F_{ij}=\epsilon_{ijk}E^k$. The fluxes then yield
\bea 
\Phi^B &=& \frac{1}{2}\int d^3x\, \,{\omega}^{ij}\,\, F_{ij} =  \frac{1}{2}\int d^3x\, \varepsilon_{ijk} \,{\omega}^{ij}\, B^{k} = \int d^3x \,{\Omega}_{k}\, B^{k}\,,\label{MagneticSmearing} \\
\Phi^E &=& \frac{1}{2}\int d^3x\,\, \tilde{\omega}^{ij}\,\,(\star F_{ij})  =  \frac{1}{2}\int d^3x\, \varepsilon_{ijk} \,\tilde{\omega}^{ij}\, E^{k} = \int d^3x \,\tilde{\Omega}_{k}\,E^{k} \,,
\label{ElectricSmearing}
\eea
where we have defined the more convenient  smearing functions
\be
\Omega_i = \frac{1}{2} \, \epsilon_{ijk} \omega^{jk}\,,\,\,\,\quad\,\,\, \tilde{\Omega}_i = \frac{1}{2} \, \epsilon^{ijk} \tilde{\omega}_{jk}\;.
\ee
This implies that the curl of $\Omega$ is given by  the current $J$ as $(\nabla \times \Omega)_i = \partial^j \omega_{ij} = J_i$. Then, for a closed curve $\Gamma=\partial\Sigma$, we have that 
\be 
\oint_{\Gamma = \partial \Sigma }  \Omega_i \, dx^i= \int_\Sigma (\nabla \times \Omega )_i \, dS^i = \int_\Sigma  J_i \,dS^i \,.
\ee
There is here an ambiguity in the choice of $\Sigma$. In particular, we are allowed to pick other surfaces like $\Sigma'$ (See figure \ref{CurveSigma}). However, because of current conservation $\nabla\cdot J=0$ and  $J$ having compact support on a ring, the flux is the same for every $\Sigma$ that cuts the ring only once. More precisely, this flux is given by the charge $q$ as
\be 
q=\int_\Sigma  J_i \, dS^i= \oint_{\Gamma}  \Omega_i\, dx^i\,.
\label{ChargeWilson2}
\ee
Note that we are using  (\ref{Condition1}) and (\ref{ChargeWilson}) to arrive at this conclusion.

\noindent
The situation is the same in the dual case with the t'Hooft loop. From (\ref{ChargeThooft}),  we can see that the monopole charge obeys
\be 
g=\int_{\tilde{\Sigma}}  \tilde{J}_i \, dS^i= \oint_{\tilde{\Gamma}}  \tilde{\Omega}_i \, dx^i  \,.
\label{chargedual}
\ee
Going back to the fluxes and using the canonical commutator between the electric and magnetic fields\footnote{This just follows from the canonical commutation relations between the electric field and the gauge potential.}
\be 
[E_i (x) , B_j (y)] =  i \varepsilon_{ijk}\partial^k \delta(x-y)\;,
\label{MaxwellCommutator}
\ee
the commutator we are seeking to compute reads
\bea
\left[\Phi^B, \Phi^E\right] &=& \int d^3x \,\, \int d^3y \,\, \tilde{\Omega}^{i}(x)\,{\Omega}^{j}(y)\,\, [B_j(y), E_{i}(x)] =  \label{CommutatorComputeM} \\
&=& i \int d^3x \,\, \int d^3y \,\, \tilde{\Omega}^{i}(x)\, (\epsilon_{ikj}\partial_y^k {\Omega}^{j}(y))\,\,  \delta(x-y) =  i \int d^3x  \,\, \tilde{\Omega}^{i}(x)\, {J}_i(x)\,. \nonumber
\eea
Inside the ring ${R}$, where  ${J}(x)$ has compact support, the other current $\tilde{J}(x)$ is zero 
\be
x\in {R}\,\Longrightarrow\, \tilde{J}(x)=0\;.
\ee
Cutting out a section ${\Sigma}$ of ${R}$ we can convert it in a simply connected region, in which we can write the vector $\tilde{\Omega}$ locally  as a gradient
\be
x\in {R}-{\Sigma}\quad \Rightarrow\quad\nabla \times \tilde{\Omega}(x) = \tilde{J}(x)= 0 \quad \Rightarrow\quad \tilde{\Omega}(x)=\nabla \tilde{\varphi}(x)\;.
\ee
The circulation of $\tilde{\Omega}$ over a non contractible curve $\gamma$ inside $R$ (linked once with $\tilde{R}$) and that includes a point $y\in {\Sigma}$ is given by the jump of $\tilde{\varphi}$ on $y$ across ${\Sigma}$. This jump is constant on ${\Sigma}$ because it is equivalent to the flux of the curl of $\tilde{\Omega}$, or the flux of $\tilde{J}$:
\be 
g= \oint_{\Gamma} \partial_i \tilde{\varphi} \, dx_i =\Delta \tilde{\varphi} (y)\;.
\label{charge2}
\ee
Using this observation inside the commutator (\ref{CommutatorComputeM}) and integrating by parts we obtain
\be
\left[\Phi^B, \Phi^E\right]= i\int d^3x  \,{J}_i(x)\,\partial^i \tilde{\varphi}(x) =  i\int d^3x  \,\,\left( \partial^i \left[\tilde{\varphi}(x) {J}_i(x)\right])- \tilde{\varphi}(x) \left[\partial^i {J}_i(x) \right]\right)\;.
\label{CommutatorCompute2}
\ee
The last term vanishes from current conservation $\partial^i {J}_i= 0$. The remaining term is a divergence that can be written as
\bea
\left[\Phi^B, \Phi^E\right]&=& i \int_{{\Sigma}} dS_i \, \Delta \tilde{\varphi} (y) \, {J}_i  =  i \, q \, g\,.
\label{CommutatorCompute3}
\eea
Physically, the charges $q$ and $g$ take into account ``how many times'' the current goes around the respective ring. Then, for thin linear regions,  this result gives the usual linking number between the linked loops. Another commentary is that due to the topological nature of these commutation relations, we can deform the regions to look like infinitely thin rings, or take the currents to be smeared over spacetime rather than simply over the $t=0$ surface, and the commutation relations do not change as far as the charges do not change and the regions remain space-like separated. Indeed, the charge preserving deformations are produced by local operators that commute with the dual non-local operator.  

If we now take the Wilson loop $W$ and the 't Hooft loop $T$ as in equations (\ref{WilsonLoop}) and (\ref{ThooftLoop}),  again with support in spatially separated rings which are however simply laced to each other, we finally arrive at the famous commutation relations \cite{tHooft:1977nqb}
\bea 
W^q\, T^g &=& e^{i\Phi^B}e^{i\Phi^E} = e^{i\Phi^B + i\Phi^E + \frac{1}{2}[\Phi^B,\Phi^E]}  = e^{ i \,q\, g}\,T^g\, W^q \,. \label{MaxwellLoops}
\eea
The subsequent terms in the Baker–Campbell–Hausdorff formula does not appear for these free fields. 

\subsection{Algebras of topological operators for the graviton}\label{GravitonCase}

In the theory of linearized gravity,  we can define analogous smeared Wilson and 't Hooft loop operators by using the conserved fluxes~(\ref{MagneticFluxes}) and~(\ref{ElectricFluxes}) related to the conserved currents (\ref{MagneticCFT}) and (\ref{ElectricCFT}). More concretely,  we define
\be
 W_G=e^{i\Phi^B_G } \,, \quad  T_F=e^{i\Phi^E_F }\,, \quad  F,\,G=P,\, J,\, D,\, K\;,\label{GravitonLoops}
\ee
where the smeared fluxes are given by 
\bea
&{}&  \Phi^B_G = \int d^3x\, \Omega^i(x) \, B^G_i (x) =\int d^3x\, \Omega^i(x) \,g^j_G(x)\, B_{ij}  (x) \\
&{}&  \Phi^E_F  =  \int d^3x\, \tilde{\Omega}^i(x) \, E^F_i (x)=\int d^3x\, \tilde{\Omega}^i(x) \,f^j_F(x)\, E_{ij}  (x)\;,
\eea
and where $g^j_G(x)$ for the different $G$ is given by
\be
 a_j\,,\,\, -s_{ j n}x^n \,,\,\, \kappa x_j \,,\,\,(b_j x^2 -2x_j b\cdot x)\,,
\ee 
while the $f^j_F(x)$ for the different $F$ are 
\be
  \tilde{a}_j\,,\,\,- \tilde{s}_{ j n}x^n \,,\,\, \tilde{\kappa} x_j \,,\,\,(\tilde{b}_j x^2 -2x_j \tilde{b}\cdot x)\,.
\ee
The smearing functions $\Omega^i(x)$ and $\tilde{\Omega}^i (x)$ are such that $\nabla \times \Omega=J$, $\nabla \times \tilde{\Omega}=\tilde{J}$. The supports of $J$ and $\tilde{J}$ are restricted to linked the ring-like regions $R$ and $\tilde{R}$ respectively. Without any loss on generality we  will set the corresponding charges to one
\be
\int_{\Sigma}  J_i \, dS^i=\int_{\tilde{\Sigma}}  \tilde{J}_i \, dS^i=1\,.
\ee

To show these are non-local operators in their respective rings, we need to prove they do not commute with certain non-local operators in the complementary region. This problem is, of course,  completely solved once we compute the algebra of the fluxes involved. 
 We start from  the expression of the commutator in terms of the smearing fields  and  consider the commutator between electric and magnetic variables
 \be
 \left[ \Phi^B_G  ,\Phi^E_F \right] =  \iint d^3x\, d^3y \, \tilde{\Omega}^{i}(x)\,  \Omega^{k} (y)\, f_F^j(x)\, g_G^l(y)\,  \left[ B_{kl} (y) , E_{ij}(x)\right] \,.
 \ee
We now replace the expression for the commutator (\ref{ElectricMagneticCommutator}), integrate by parts the  derivatives of each term acting over $y$, and integrate out the delta function by integrating over $y$.  As a result,
we get that the commutator reads
\begin{align}
& \left[ \Phi^B_G  ,\Phi^E_F \right]  = -\frac{i}{4}   \int d^3x\,  \tilde{\Omega}^{i} f_F^j \, \left[ (\partial_j \partial_l - \delta_{jl} \partial^2)J_{i}+  (\partial_i \partial_l - \delta_{il} \partial^2) J_{j}  -(\partial_i \partial_j- \delta_{ij} \partial^2) J_{l}  \right]g_G^l \nonumber \\
&-\frac{i}{4}   \int d^3x\,  \tilde{\Omega}^{i} f_F^j \, \left[ \varepsilon_{ibk}(\partial_j \partial_l - \delta_{jl} \partial^2)+   \varepsilon_{jbk}(\partial_i \partial_l - \delta_{il} \partial^2) - \varepsilon_{lbk}(\partial_i \partial_j- \delta_{ij} \partial^2)  \right]\Omega^k \partial_b g_G^l\;,\label{CommutatorCompute}
 \end{align} 
 where we have used $J_i= \varepsilon_{ijk}\,\partial^j \Omega^k$. 

The second term on the right-hand side of (\ref{CommutatorCompute}) is not expressed in terms of the current. This momentarily obstructs the localization of the integral in the region $R$ (which is the support of $J$). This localization was fundamental in the derivation of the commutation relations for the Maxwell field. However, this localization occurs when we write in detail the form of the function $g_G^l$.
We have to analyze case by case. We provide here the example for ``magnetic dilatations''  $g^l_D(x) = \kappa\, x^l $. We get
\bea
&{}&  \left[ \varepsilon_{ibk}(\partial_j \partial_l - \delta_{jl} \partial^2)+   \varepsilon_{jbk}(\partial_i \partial_l - \delta_{il} \partial^2) - \varepsilon_{lbk}(\partial_i \partial_j- \delta_{ij} \partial^2)  \right]\left(\Omega^k \partial^b g^l\right) =\label{SecondTerm} \\ 
&{}&  =\kappa \left[ \varepsilon_{ilk} \partial_j \partial^l   \Omega^k -  \varepsilon_{ijk} \partial^2  \Omega^k+   \varepsilon_{jlk}\partial_i \partial^l   \Omega^k -  \varepsilon_{jik} \partial^2  \Omega^k\right] =  \kappa \left[ \partial_j    J_i +  \partial_i J_j \right]\;.  \nonumber 
\eea
Then, it can finally be expressed in terms of the current as happens for the first term in (\ref{CommutatorCompute}).
Substituting into the expression for the commutator,  we find
\bea
  \left[ \Phi^B_D  ,\Phi^E_F \right]  = - \frac{i\, \kappa}{4}   \int d^3x\,  &\tilde{\Omega}^{i} f_F^j &\, \left[ (\partial_j \partial_l - \delta_{jl} \partial^2) (J_{i} x^l)+  (\partial_i \partial_l - \delta_{il} \partial^2) (J_{j} x^l ) \right. \nonumber\\
 & {} & \left. -(\partial_i \partial_j- \delta_{ij} \partial^2) (J_{l} x^l)+\partial_j    J_i +  \partial_i J_j  \right]\,. \label{d1}
 \eea
Since every term is proportional to the current, the integration restricts to the region $R$.
 In (\ref{d1}) we are considering the particular case of $g=g_D$, but this step happens for the other $g_G$ as well.  In the other cases,  the analog of (\ref{d1}) is given for``magnetic translations, rotations,  and special conformal transformations'' by
\begin{align}
&  \left[ \Phi^B_P  ,\Phi^E_F \right] = -\frac{i\, \kappa}{4}   \int d^3x\,   \tilde{\Omega}^{i} f_F^j \, \left[ (\partial_j \partial_l - \delta_{jl} \partial^2) (J_{i} x^l)+  (\partial_i \partial_l - \delta_{il} \partial^2) (J_{j} x^l ) \right. \nonumber\\
 &\qquad\qquad\qquad\qquad\qquad \left. -(\partial_i \partial_j- \delta_{ij} \partial^2) (J_{l} x^l) \right]\,, \label{d1p} \\
& \left[ \Phi^B_J  ,\Phi^E_F \right] = -\frac{i\,s^{nl}}{4}   \int d^3x\,  \tilde{\Omega}^{i} f_F^j \, \left[ (\partial_j \partial_l - \delta_{jl} \partial^2) (J_{i} x^n)+  (\partial_i \partial_l - \delta_{il} \partial^2) (J_{j} x^n ) \right. \nonumber\\
 &\qquad\qquad\qquad\qquad  \left. -(\partial_i \partial_j- \delta_{ij} \partial^2) (J_{l} x^n)+\frac{1}{2}\left(\varepsilon_{inl}\varepsilon_{jab}+\varepsilon_{jnl}\varepsilon_{iab}\right)\partial^a J^b \right]\,, \label{d1j}\\
 &  \left[ \Phi^B_K  ,\Phi^E_F \right] = -\frac{i\, }{4}   \int d^3x\,  \tilde{\Omega}^{i} f_F^j \, \left[ (\partial_j \partial_l - \delta_{jl} \partial^2) (J_{i} x^l)+  (\partial_i \partial_l - \delta_{il} \partial^2) (J_{j} x^l ) \right. \nonumber\\
 &\qquad\qquad\qquad\qquad \left. -(\partial_i \partial_j- \delta_{ij} \partial^2) (J_{l} x^l)+2(b\cdot x)\left(\partial_j    J_i +  \partial_i J_j\right)-8 \,\delta_{ij}\, (b\cdot J)  \right. \label{d1k} \\
  &\qquad\qquad\qquad\qquad  \left. + 6\, b_i \,J_j+ 6\, b_j\, J_i+(b^l\, x^n - b^n \,x^l ) \left(\varepsilon_{inl}\varepsilon_{jab}+\varepsilon_{jnl}\varepsilon_{iab}\right)\partial^a J^b \right]\,. \nonumber
 \end{align}

  The next step is to write in $R$ the field $ \tilde{\Omega}_{i} = \partial_i \tilde{\varphi}$ where $\tilde{\varphi}$ has a unit jump across $\Sigma$. After integration by parts we find that the divergence in the index $i$ of the integrand vanishes. This holds for all  $f_F (x)$, $F=P,D,J,K$, and $g_G (x)$, $G=P,D,J,K$ considering current conservation. For example, for  the case of magnetic dilatations, eq. (\ref{d1}), 
 \begin{align}
 \partial_i  \left\{ f_F^j \, \left[ (\partial_j \partial_l - \delta_{jl} \partial^2)\right.\right. & (J_{i} x^l)+  (\partial_i \partial_l - \delta_{il} \partial^2) (J_{j} x^l )  \nonumber\\
 & \left.\left. -(\partial_i \partial_j- \delta_{ij} \partial^2) (J_{l} x^l)+\partial_j    J_i +  \partial_i J_j  \right]\right\}=0\,. \label{div}
 \end{align}
As a result of (\ref{div}),  the integral in (\ref{d1}) is given by the surface boundary term, which is independent of the particular surface $\Sigma$ that cuts $R$:
 \bea 
  \left[ \Phi^B_D  ,\Phi^E_F \right] = - \frac{i\, \kappa }{4}\, \int_\Sigma &dS_i&\,    f^F_j \left[(\partial_j \partial_l - \delta_{jl} \partial^2) (J_{i}x_l )+ (\partial_i \partial_l - \delta_{il} \partial^2)  (J_{j}x_l )\right. \nonumber \\   
   &{}&\left.-(\partial_i \partial_j- \delta_{ij} \partial^2)  (J_{l}x_l )+\partial_j    J_i +  \partial_i J_j \right] \;.
 \eea

\begin{figure}[t]
\includegraphics[width=8cm]{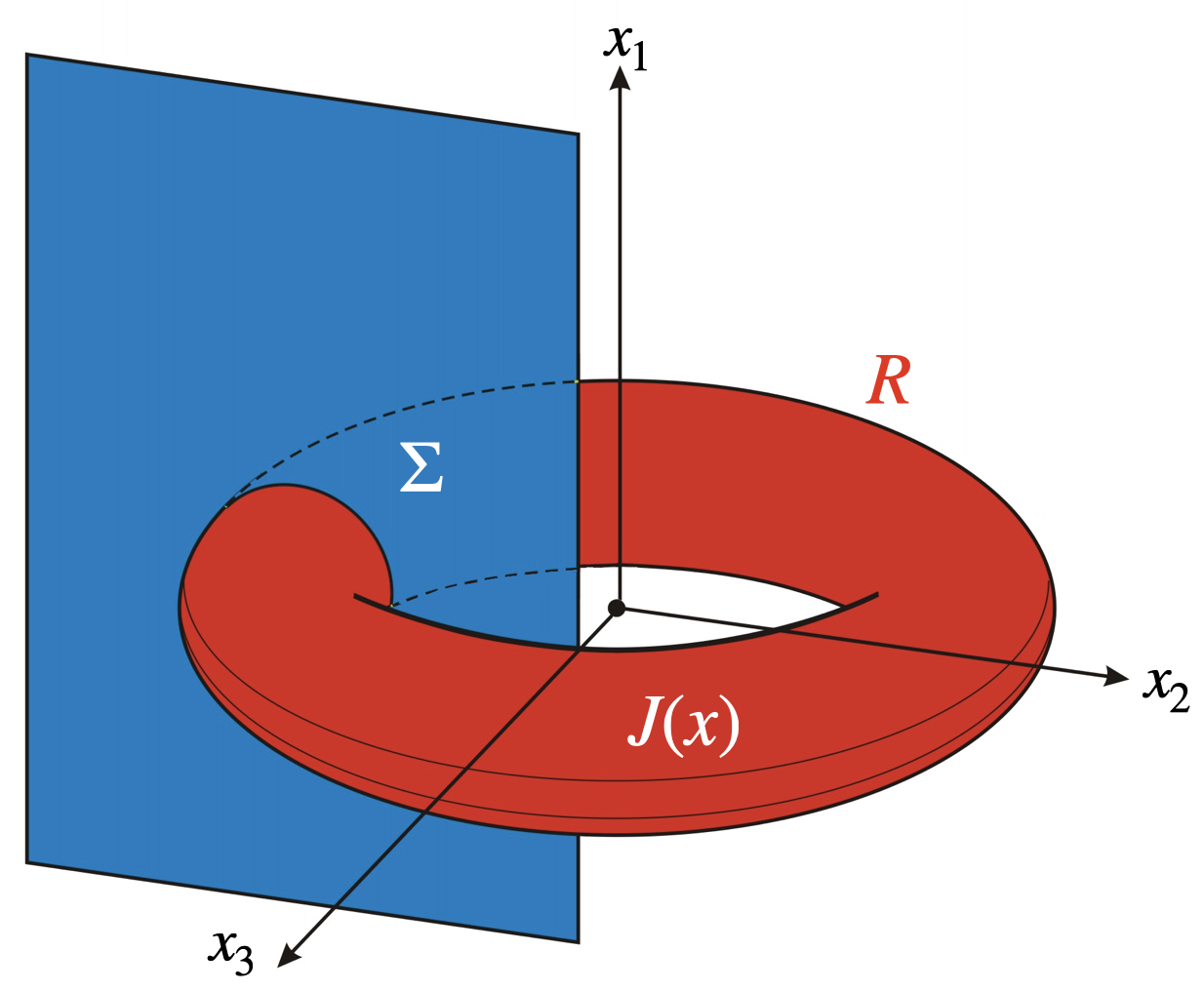}
\centering
\caption{Example of the integration surface $\Sigma$. Here we use coordinates $(x_1,x_2,x_3)$ with the surface $\Sigma$ being orthogonal to $x_3$. }
\label{halfplane}
\end{figure}

To evaluate the surface integral, considering that the result is independent of the chosen cross-section of $R$, we choose a planar surface $\Sigma$. We take Cartesian coordinates $x_1,x_2$ on this surface, and $x_3$ is the orthogonal one. See figure \ref{halfplane}.
With this picture in mind, let's study the surface integrals that appear. These are of the form
  \bea 
 \left[ \Phi^B_D  ,\Phi^E_F \right] =  -\frac{i\, \kappa }{4}\, \int_\Sigma dx_1\, dx_2\, &&    f_F^j \left[(\partial_j \partial_l - \delta_{jl} \partial^2) (J_{3}\,x^l )+ (\partial_3 \partial_l - \delta_{3l} \partial^2)  (J_{j}x^l )\right. \nonumber \\   
   &{}&\left.-(\partial_3 \partial_j- \delta_{3j} \partial^2)  (J_{l}x^l )+\partial_j    J_3 +  \partial_3 J_j \right] \;.
 \eea
The integration of derivatives parallel to the surface vanishes because the current and its derivatives have compact support in $R$ and vanish on the boundaries of $\Sigma$. We can also use current conservation to write $\partial_3 J_3=-\partial_1 J_1-\partial_2 J_2$, and integrate by parts the parallel derivatives. After some algebra,  we find that for any specific $f_F^j$ the integral, due to quite remarkable and convenient cancellations, either vanishes or gets proportional to $\int_\Sigma dx_1\,dx_2\,J_3=1$.   
In this way we get
     \be
 \left[ \Phi^B_D  ,\Phi^E_P \right] =0 \,, \quad   \left[ \Phi^B_D  ,\Phi^E_J \right]=0\,, \quad   \left[ \Phi^B_D  ,\Phi^E_D \right] =  \frac{i\, \kappa\, \tilde{\kappa} }{2}\,, \quad   \left[ \Phi^B_D  ,\Phi^E_K \right] = 0\;.
 \ee

As mentioned, the same procedure can be followed to obtain the other commutators for $g_G$ with $G\neq D$. We are not reproducing these calculations here because they are quite long and not particularly enlightening. We have also checked the result performing the calculations with a program for mathematical manipulations. The complete algebra of topological operators that we obtain considering (\ref{d1}), (\ref{d1p}), (\ref{d1j}) and (\ref{d1k}) is the following
\bea
&{}&   \left[ \Phi^B_P  ,\Phi^E_K \right]  = i\,  a^i\,\tilde{b}_i  \,, \quad   \left[ \Phi^B_J  ,\Phi^E_J \right]=\frac{i}{4} \,s^{ij}\,\tilde{s}_{ji} \,,\nonumber \\ &{}&  \left[ \Phi^B_D  ,\Phi^E_D \right]=\frac{i}{2}\, \kappa\,\tilde{\kappa} \,, \quad \left[ \Phi^B_K  ,\Phi^E_P \right]=i\,b^i\, \tilde{a}_i  \,.\label{ccc}
\eea
Note that all the other commutators have to be zero for dimensional reasons. The commutators are dimensionless as well as the charges. To get a dimensionless result that depends entirely on the scalar, vector, and tensor charges the only commutators that can exhibit non-zero results are the above ones. Other possibilities involve functions of the coordinates depending on the geometry of the rings, but this is not possible due to the topological nature of these charges. If we smoothly change the geometry of $R$ and the value of $J$  to $R_1$ and $J_1$, while keeping the charge fixed and avoiding intersecting $\tilde{R}$,  the commutators must not change. The reason is that the difference between the flux operator defined at $R$ and the one defined at $R_1$ is an additive operator in the complement of $\tilde{R}$, commuting with operators based at $\tilde{R}$.  

The result (\ref{ccc}) can be further checked using singular (infinitely thin) linear loops. In the appendix app~(\ref{Squares}),  we show how to compute them using square loops.

In consideration of these results, the corresponding non-zero commutation relations for the graviton Wilson and 't Hooft   loops (\ref{GravitonLoops}) are given by 
\bea
&{}& W_P^a \,T_K^{\tilde{b}} = e^{i \tilde{b}\cdot a } \,T_K^{\tilde{b}} \, W_P^a  \,, \quad\,\,\,   W_J^s \,T_J^{\tilde{s}} = e^{-\frac{i}{4} \tilde{s}\cdot\cdot s } \, T_J^{\tilde{s}} \, W_J^s \,,\nonumber \\ &{}&  W_D^\kappa \,T_D^{\tilde{\kappa}} = e^{\frac{i}{2}\,\tilde{\kappa} \kappa } \,T_D^{\tilde{\kappa}} \, W_D^\kappa  \,, \quad  \,W_K^b \,T_P^{\tilde{a}} = e^{i \tilde{a}\cdot b } \, T_P^{\tilde{a}} \, W_K^b  \,.
\eea
The commutation relations show that the previously defined operators are non-local operators in the ring, since they do not commute with at least one non-local operator in the complementary ring. The group of generalized symmetries is the Abelian group $\mathbb{R}^{10}\times (\mathbb{R}^{10})^*$ of electric and magnetic fluxes. There are $10$ independent electric and $10$ independent magnetic fluxes, both of which can be based on the same ring-like region $R$ in $d=4$. 

The most salient feature of this generalized symmetry, which is not present for the ordinary gauge theories, is the presence of Lorentz indices for the generalized symmetry charges. We will make further remarks on this issue in the discussion section. The sectors with vector indices may be an oddity in relativistic theories, but this is not so for condensed matter systems with fracton excitations, as we now describe.

\section{Fractons and graviton completeness}\label{FR}

Having derived the algebra of generalized symmetries of the free graviton theory, we now discuss some interesting avenues and applications. In particular, as described in the introduction, we want to approach the issue of charging or breaking these generalized symmetries from a purely QFT perspective. 

Let us be more precise. In \cite{Polchinski:2003bq} it was argued that gauge effective field theories in quantum gravity should be ``complete'', meaning the spectrum of possible charges should be maximal consistent with the Dirac quantization condition. This principle was extended later in \cite{Banks:2010zn} to include other generalized symmetries. More recently in \cite{Rudelius:2020orz,Casini:2020rgj,Heidenreich:2020tzg}, the relation between completeness and the absence of generalized symmetries has been further developed and deepened. In particular, Ref. \cite{Casini:2020rgj} defines a QFT (without gravity) to be complete whenever relation~(\ref{caus}) is saturated for the observable algebra and any region with any given topology. Allowing for non-saturation of this condition inevitably leads to the existence of two dual generalized symmetries, as the two dual sets we derived for the graviton.\footnote{In the context of AdS/CFT there are different arguments for the absence of different types of symmetries in bulk effective field theories, see \cite{Harlow:2018tng,Harlow:2015lma,Harlow:2021trr,Review}.} From the existence of such generalized symmetries, or equivalently from the non-saturation of~(\ref{caus}) in ring-like regions, we know the free graviton theory is not complete. 

It is thus interesting to see whether we can ``complete'' the graviton theory by introducing a sufficient number of charged operators to destroy the set of generalized symmetries described above. One might anticipate that problems should appear when attempting to carry out this process. For example, since the generalized symmetries of the graviton are charged under space-time symmetries, we expect some generalization of the Coleman-Mandula theorem to generalized internal symmetries to make it difficult to find a simple Lorentz invariant completion of this theory. Also, from the present knowledge on quantum gravity, we do not expect to complete the theory in a nice relativistic QFT fashion.

It turns out that the fact that the generalized symmetries of the graviton are charged under space-time symmetries, coming from the fact that the conserved fluxes are themselves charged under these symmetries, means that the charged particles making the topological operators additive, will be highly constrained since their movement in spacetime has to respect all those conservation laws. Quite interestingly, this type of behavior has been observed very recently in the context of condensed matter systems, where it comes under the name of ``fractons''.

 \subsection{Fractons and tensor gauge theories}
 
Fractons are defined to be particles with inability to move through space \cite{OrginalFracton1,OrginalFracton2,OrginalFracton3,OrginalFracton4,Pretko:20161}.\footnote{In this line one also finds ``lineons'' and ``planons'', which are particles allowed only to move in appropriate submanifolds \cite{Pretko:2020}.} The reason has nothing to do with inertia, in the sense of having large masses. It has to do with quite peculiar conservation laws that their motion has to respect.  Although fractons, in isolation, cannot move, bound states of fractons can move through space. For example, in some models to be reviewed below, single fractons cannot move while dipoles can. Intuitively, there is a ``symmetry'' that prohibits the existence of localized dipole operators in the theory, which otherwise could that transport fractons from one place to another, but that symmetry does not prohibit the existence of localized quadrupole operators transporting dipoles from place to place, see the recent review \cite{Pretko:2020} and references therein, and also \cite{exp1,exp2} for potential experimental senarios.

These quite unusual and striking properties have been nicely described by Pretko \cite{Pretko:20161,Pretko:20162} in terms of recent advances in tensor gauge theories \cite{Rasmussen2016}, which happen to display higher moments conservation laws. A standard example one finds in the literature is the following. We can start with a two index symmetric gauge potential $A_{ij}$ and a gauge transformation law given by
\be \label{gauge}
A_{ij}\rightarrow A_{ij}+\partial_i\partial_j \alpha\;.
\ee
A two index gauge invariant magnetic field can be defined as usual by
\be 
B_{ij}=\epsilon_{ikl}\partial^k A^l_j\;,
\ee
and the gauge invariant electric field can be defined implicitly by imposing canonical commutation relation with the gauge potential
\be 
[A_{ij},E_{kl}]=i(\delta_{ik}\delta_{jl}-\delta_{il}\delta_{jk})\delta(x-y)\;.
\ee
The constrain equation of this theory is a modified version of the Maxwell case
\be 
\partial^i\partial^j E_{ij}=0\;,
\ee  
which one can verify generates the gauge transformation (\ref{gauge}) given the canonical commutation relations. Introducing a new potential $A_0$ with gauge transformation 
\be A_0\rightarrow A_0+\dot{\alpha}\,,
\label{gaugeg}
\ee
 the electric field can be written $E_{ij}=\dot{A}_{ij}-\partial_i \partial_j A_0$.

 It is of interest to understand how to couple this model with sources. Calling $\rho=\partial^i\partial^j E_{ij} $ to a charged density source of the previous constraint equation, the electric conserved charges of this model can be measured by the flux at infinity
\be 
Q=\int dx^3 \rho=\int dx^3\partial_i\partial_j E_{ij} =\int dS_j \partial_i E^{ij}\;,
\ee
in the same way as one does with the usual electric charge. In addition, we also have a vector charge, non invariant  under the space-time symmetries. This ``dipole'' charge writes
\be 
P^i=\int dx^3 \rho \,x^i=\int dx^3 x^i\partial_j\partial_k E^{jk} =\int dS_j (x^i\partial_k E^{jk}- E^{ij})\;.
\ee
The conservation of this dipole charge impedes local charged excitations, called fractons, to move around space, while dipoles can move if they conserve the direction. This model is called the ``scalar charge tensor model''.

A second type, the ``vector charge tensor model'', arises by considering the following different gauge transformation for the potential
\be \label{gauge2}
A_{ij}\rightarrow A_{ij}+\partial_i \alpha_j+\partial_j \alpha_i\;,
\ee
generated by the following vector constraint
\be 
\partial^i E_{ij}=0\;.
\ee  
The gauge invariant magnetic field can be written as
\be 
B_{ij}  =\varepsilon_{iab}\varepsilon_{jcd} \partial^b \partial^d A^{ac}\,,\label{410}
\ee
which gives $\partial^i B_{ij}=0$.

This is called a vector theory since the charged source of the constraint transforms like a vector under rotations. This theory has the usual conserved charge
\be 
 Q^i=\int d^3x\, \rho^i = \int d^3x\,\partial_j E^{ij}=\int dS_j\, E^{ij}\;,
\ee
  but also an angular momentum type charge
\be 
  M^i=\int  d^3x\,\varepsilon^{ijk} \rho_j x_k =\int  dS_j\, (\varepsilon^{ilk} \,E^{j}_{\,\,l}\,x_k )\;.
\ee

Again, the conservation of both charges constraints the movement of excitations and bound states. In general, these models are included inside ``higher rank gauge theories'', whose degrees of freedom are gauge potentials and electric and magnetic fields with several indices. Constraints arise by different combinations of divergences and traces, see \cite{Rasmussen2016} for a nice and short account.
  
Given these features of the tensor gauge theories that potentially couple to fractons or related excitations, connections with general relativity were developed in \cite{Gu2006,Xu2006,Pretko:2017}. In particular, it was argued in \cite{Pretko:2017} that several properties of the interactions between fractons resemble gravity. The heuristic reasons described in those papers were rooted in the fact that the graviton theory, in its metric formulation, can be seen as a short of tensor gauge theory for the gauge potential $h_{\mu\nu}$, the deviation from the background metric. The role of the non-trivial conservation laws of the tensor gauge theories was argued to be played by the Hamiltonian and momentum constraints (\ref{Constraints}).

Instead of trying to adapt the zoo of fractonic systems and tensor gauge theories to gravity, using our previous results on generalized symmetries and the generalized electromagnetic formulation of linearized gravity we can go in the opposite direction. We can describe the linearized graviton theory as a specific example of the zoo of gauge theories describing fractonic systems.

Looking at the previous classification and our results above, we see that the constraints on the physical variables $E_{ij}, B_{ij}$ inform us that the graviton is an example of a ``traceless vector charge theory'', as described in \cite{Pretko:20162}. Interestingly, the graviton theory is not precisely the higher tensor gauge theory of this type described in \cite{Pretko:20162}, but it shares with it the same set of conservation laws and generalized symmetries, the reason being that the model is also self-dual, and the electric and magnetic fields satisfy the same algebraic and conservation relations as stress tensors of euclidean field theories in one less dimension. The difference lies in the dynamics and commutation relations. In particular, the model described in \cite{Pretko:20162} must have the same structure of non-local operators and generalized symmetries as the ones described in the present paper. However,  we expect the algebra of these topological operators to be different from the algebra described above, coming from the graviton theory.

 Also, the natural way to couple fractonic matter to gauge fields does not turn out to be the same as the way we couple matter to the metric in general relativity. In particular, we can write the linearized equations of motion in the presence of a minimally coupled matter stress tensor as
\be 
G_{\mu\nu}^{(1)}=R_{\mu\nu}^{(1)}-\frac{\eta_{\mu\nu}}{2}R^{(1)} =  T_{\mu\nu}\,,
\ee
where $T_{\mu\nu}$ accounts for matter and graviton sources as in \cite{Weinberg:1972kfs}. In the electromagnetic formulation, this is
\begin{align}
&E_i^i = T^i_i-2T_{00}\,,\,\quad \quad \quad \quad\quad \quad \quad \epsilon_{ijk}E_{ij} =0\,\nonumber,\\
&B_i^i = 0\,,\,\,\,\quad \quad \quad\,\,\quad \quad \quad \quad \quad  \quad \quad  \epsilon_{ijk}B_{ij} =T_{k0}\,, \\
&\partial_j E_{ij}=\dot{T}_{0i}-2\partial_i T_{00}+\partial_i T_j^j\,,\quad \quad  \epsilon_{iab}\partial_a E_{bj}=-B_{ij}\,,\nonumber \\
&\partial_j B_{ij}=\epsilon_{ijk}\partial_j T_{k0}\,,\,\,\,\quad\quad\quad  \quad \quad \quad   \epsilon_{iab}\partial_a B_{bj}=E_{ij}-T_{ij}-\frac{\delta_{ij}}{2}T+\partial_j T_{ij}\,.\nonumber
\end{align}
The equivalence principle, in its ``minimal coupling'' incarnation, forces the ``electric and magnetic currents'' to have some definite peculiar form. This feature also deviates from the specific model considered in \cite{Pretko:20162}.

 \subsection{Wilson lines for the graviton}

If we were considering a conventional gauge theory, to destroy the generalized symmetry we would first write the non-local operator as an additive operator by using the gauge potential. In other words, by writing the non-local operator as a loop operator of a non-gauge invariant field, instead of writing it as a surface flux. 
By doing so we can now break the loop into open Wilson lines. These open Wilson lines are not gauge invariant, but their transformation is only dependent on the endpoints. This tells us what kind of matter we should include to obtain a truly gauge invariant Wilson line. For typical gauge theories, this means to include matter transforming under a certain representation of the gauge group. These gauge invariant Wilson lines then convert the topological operator into an additive one.  

We have only produced an expression of the non-local operators in terms of gauge invariant fluxes. We now find expressions in terms of loop operators of non-gauge invariant fields, playing the role of the gauge potential in the conventional gauge theory scenario. These expressions suggest the matter content that can break the non-local operators is a nonconventional completion of the linearized graviton theory.
  
To start with, we notice the  electric and magnetic fields of the graviton can be written in terms of the extrinsic curvature (\ref{ExtrinsicCurvature}) and lapse function (\ref{Foliation}) as
\be
E_{ij}= \dot{K}_{ij} - \partial_i  \partial_j N\,, \quad
B_{ij}= \varepsilon_{iab}\partial_b K_{aj}\,.
\label{ElectricMagneticCurvature}
\ee
Although the electric and magnetic fields are invariant under the action of linearized diffeomorphisms, the extrinsic curvature and lapse function transform as a ``scalar charge tensor model'' with the law (\ref{gauge}) associated to the component $\xi_0$ as
\be 
K_{ij}\to K_{ij}-\partial_i \partial_j\xi_0\,, \quad N\to N - \dot{\xi}_0\,.
\label{gaugeKN}
\ee
This only depends on the function  $\xi_0$ and reproduces the transformations (\ref{gauge}) and (\ref{gaugeg}) presented in \cite{Pretko:20161,Pretko:20162}. Notice however that $E_{ij}$ and $K_{ij}$ are not canonically conjugated variables, as usually assumed in the discussions of scalar charge tensor models. At this point, the similitude of the graviton theory with these types of models disappears.  In particular, the transformation of the canonical variables induced by the constraints is very different. 

\begin{figure}[t]
\centering
\includegraphics[width=.5\textwidth]{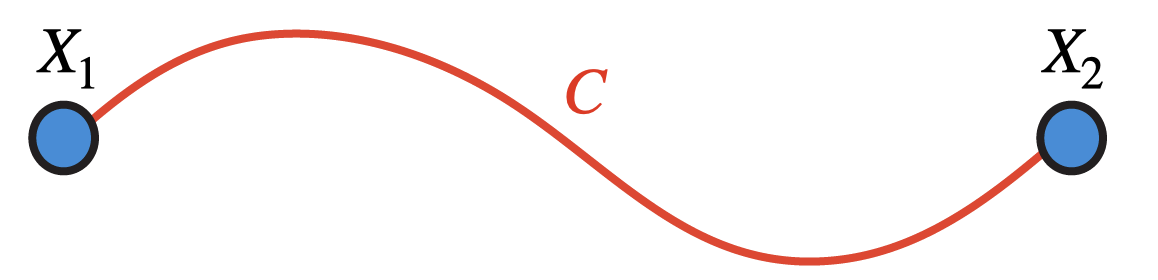}
\caption{Setup of a curve that defines a line operator}
\label{LineOperator}
\end{figure}

Using the expressions (\ref{ElectricMagneticCurvature}) and (\ref{gaugeKN})  we can write the traslation and dilatation magnetic fields described in the equation (\ref{MagneticCFT}) as 
\be
B_i^P= \epsilon_{iab}\partial^b ({K}^{aj}a_j)\,, \quad B_i^D= \epsilon_{iab}\partial^b (\kappa \,{K}^{aj}x_j)\;.
\ee
This allows us the write the corresponding non-local operator fluxes as circulations of ${K}^{aj}a_j$ and $\kappa {K}^{aj}x_j$ along the boundary of the surface.  These same line integrals over an open curve $C$ define   
\begin{align}
U^P(C)= &\int_C d\ell_i \, ({K}^{ij}\,a_j) =  \frac{1}{2}\int_C d\ell_i \, (a_j \dot{h}^{ij} - a_j \partial^j h_{0}^{\,\,i}- a_j \partial^i h_{0}^{\,\,j}) \,, \\ 
 U^D(C)= &\int_C d\ell_i \, (\kappa\,{K}^{ij}\,x_j)  =  \frac{\kappa}{2}\int_C d\ell_i \, (x_j \dot{h}^{ij} - x_j \partial^j h_{0}^{\,\,i}- x_j \partial^i h_{0}^{\,\,j})\,.
\end{align} 
These, as expected, have gauge transformations that only depend on the endpoints of the curve (see figure \ref{LineOperator})
\begin{align}
U^P(C) \to  & \,\,\, U^P(C) - a_j \int_C d\ell_i \,\partial_i \partial_j \xi_0 =   U^P(C) -\left. a_j \partial_j \xi_0   \right|^{X_2}_{X_1}\,,
\label{g1} \\
U^D(C) \to &   \,\,\,U^D(C) - \kappa \int_C d\ell_i \, x_j \partial_i \partial_j \xi_0 =   U^D(C) -  \kappa \left.\left( x_j \partial_j \xi_0 - \xi_0  \right)\right|^{X_2}_{X_1}\,.
\label{g2}
\end{align} 

\begin{figure}[t]
\centering
\includegraphics[width=0.8\textwidth]{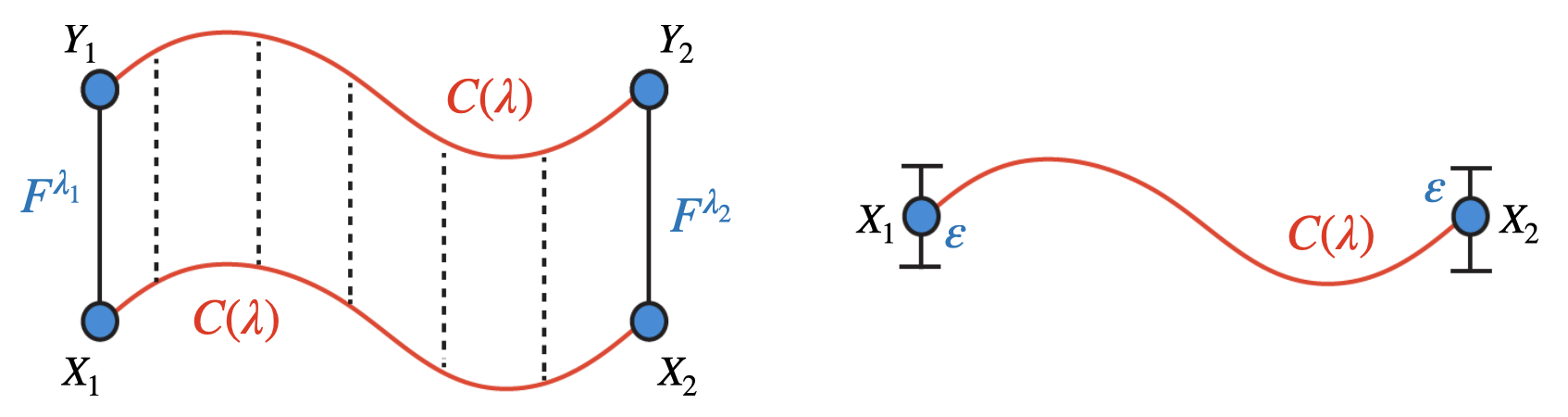}
\caption{Setup of the curves corresponding to a finite Wilson strip (left) and the associated infinitesimal limit or line operator (right).}
\label{c2}
\end{figure}

We can interpret these Wilson lines as the limit of Wilson strips associated with a fracton dipole, as it is done in \cite{Seiberg:2020wsg,Seiberg:2020cxy}  in the context of scalar charges tensor models. The corresponding setup is described in Figure \ref{c2}. The strip operator is now given by a double line integral over the curve $C$ and the fiber $F_\lambda$ at each point $\lambda$ of $C$: 
\be 
U(C,F)=\int_{C}d\ell^C_i \int_{F_\lambda}d\ell^\lambda_j \, K^{ij} = \int_{\lambda_1}^{\lambda_2}d\lambda \int_{\beta_1}^{\beta_2}d\beta \, \left[\frac{\partial x^C_i (\lambda)}{\partial \lambda}\frac{\partial x^\lambda_j (\beta)}{\partial \beta}K^{ij}\right]\,.
\label{StripOperator}
\ee
For an infinitesimal fiber we write the tangent vector $\frac{\partial x^\lambda_j (\beta)}{\partial \beta}=\epsilon \, k^j (\lambda)$ and we get 
\be 
 U(C,F)= \epsilon \int_{\lambda_1}^{\lambda_2}d\lambda  \, \frac{\partial x^C_i (\lambda)}{\partial \lambda}k_j(\lambda) \, K^{ij}\,.
 \label{wl}
 \ee
From here, gauge transformations (\ref{gaugeKN}) act as 
 \begin{align}
 \delta U(C,F)= & - \epsilon \int_{\lambda_1}^{\lambda_2}d\lambda  \, \frac{\partial x^C_i (\lambda)}{\partial \lambda}\,k_j(\lambda) \,\partial_i\partial_j \xi_0 = - \epsilon \int_{\lambda_1}^{\lambda_2}d\lambda  \, k_j(\lambda) \, \partial_\lambda\partial_j \xi_0=\nonumber \\
 &  =- \epsilon  \left(\left. k_j(\lambda)\,\partial_j \xi_0\right) \right|^{\lambda_2}_{\lambda_1} + \epsilon \int_{\lambda_1}^{\lambda_2}d\lambda  \, \left[ \frac{\partial k_j (\lambda)}{\partial \lambda}\partial_j \xi_0 \right]\,.   \label{gt}
 \end{align}
This has a term that is not dependent exclusively on the endpoints of the curve $C(\lambda)$. However, we are still able to choose the dependence of the fiber direction as we want. To reproduce the gauge transformations of the translations line operator $U^P(C)$ we choose 
$k_j(\lambda)= a_j$. In this way,  we recover 
\be
U(C,F)=\epsilon \int_{\lambda_1}^{\lambda_2}d\lambda  \, \frac{\partial x^C_i (\lambda)}{\partial \lambda}\, a_j\,K^{ij} =  \epsilon  \int_C d\ell_i \,({a}_j\,K^{ij})=\epsilon\, U^P(C)\,,
\ee
having a gauge transformation consistent with the expected (\ref{g1})
\be 
 \delta U(C,F) = - \epsilon\, \left.\left({a}^j\partial_j \xi_0\right) \right|^{X_2}_{X_1}\,.
 \ee
\noindent
Considering dilatations we  propose the dependence  
$
k_j(\lambda)= \kappa\, x^C_j(\lambda)
$ 
and get 
\be 
U(C,F)=\epsilon \, \kappa\,\int_{\lambda_1}^{\lambda_2}d\lambda  \, \left[\frac{\partial x^C_i (\lambda)}{\partial \lambda}\,  x^C_j(\lambda)\,K^{ij}\right]  = \epsilon\, \kappa\,  \int_C d\ell^C_i ( x^C_j\,K^{ij})=\epsilon \,U^D(C)\,. 
 \ee
With this choice of fiber, the second term in (\ref{gt}) gives a boundary term, and we recover the expected gauge transformation from (\ref{g2})
 \be 
  \delta U(C,F)=- \epsilon\,\kappa\, \left.\left( x^C_j (\lambda) \partial^j \xi_0\right) \right|^{\lambda_2}_{\lambda_1} +\epsilon\, \kappa\, \int_{\lambda_1}^{\lambda_2}d\lambda  \, \left[ \frac{\partial x^C_j (\lambda)}{\partial\lambda}\partial^j \xi_0 \right] =   - \kappa \,\epsilon \,\left.\left( x_j  \partial^j \xi_0 - \xi_0\right) \right|^{X_2}_{X_1} \,.
 \label{acc}\ee
In this way, it is evident that one form to make these Wilson lines gauge invariant is to couple the exponentials of these strip operators with ordinary charge dipoles at the extremes, where the charged field transforms, as usual, as $\psi\rightarrow e^{i \xi_0} \psi $. We need a constant dipole for the translational charge and a dipole of size proportional to $\epsilon x_i$ for the dilatation charge. In this last case, we need also a non-zero charge according to (\ref{acc}).    Some candidates for the fractonic matter Lagrangians that obey these exotic transformation laws have been recently studied in \cite{Bidussi:2021}.
 
More complex arrangements are necessary for the other charges. We consider the case of rotation charges. To our knowledge, there is no way to write the Wilson lines related to the rotations operators coming from (\ref{MagneticCFT}) as loop integrals of non-gauge invariant operators. We can go further by looking at the dual theory presented in section (\ref{DualSection}). Here, we consider the electromagnetic formulation from (\ref{DualityB}) and (\ref{DualityE}), that is,
\be
E_{ij} =-\frac{1}{2}\varepsilon_{iab}\varepsilon_{jcd} \partial^b \partial^d {h}^{ac}\,,\quad B_{ij}  =\frac{1}{2}\varepsilon_{iab}\varepsilon_{jcd} \partial^b \partial^d \tilde{h}^{ac}\,.
\ee 
Note that now the gauge symmetry has the aspect of a ``vector charge tensor model'' as in (\ref{gauge2}) and (\ref{410}) (compare with \cite{Rasmussen2016}). This is given by 
\be 
 \tilde{h}_{ij}\to \tilde{h}_{ij}+\partial_i \tilde{\xi}_j +\partial_j \tilde{\xi}_i\,.
\ee 
Then, we write the translations and rotations currents in (\ref{MagneticCFT}) as
\be
B_i^P =\frac{a^j}{2}\varepsilon_{iab}\varepsilon_{jcd} \partial^b \partial^d \tilde{h}^{ac}\,, \quad B_i^J= \frac{s^{jn}}{2}\, x_n\, \varepsilon_{iab}\varepsilon_{jcd} \partial^b \partial^d \tilde{h}^{ac}\,, \quad s_{ij}=\varepsilon_{ijk}s^k\;.
\ee
The corresponding Wilson lines are
\begin{align}
U^P(C)= & \frac{1}{2}\int_C d\ell_i \, \left[ \varepsilon_{jab}\,\partial^b \tilde{h}^{ai}\,a^j \right] \,,  \\ 
 U^J(C)= & \frac{1}{2}\int_C d\ell_i \, \left[ (s_n x_k - s_k x_n )\, \partial^n \tilde{h}^{ki}- s_k\,\tilde{h}^{ki}  \right]\,. 
\end{align}
Now, as wanted, the corresponding gauge transformations only depend on the endpoints of the curve:
\begin{align}
U^P(C) \to  & \,\,\,   U^P(C) -\left. \frac{1}{2}\varepsilon^{jab} a_j \partial_b \tilde{\xi}_a   \right|^{X_2}_{X_1}\,,
\label{g11} \\
U^D(C) \to &  \,\,\, U^D(C) -  \frac{1}{2} \left.\left(s^n x^k -s^k x^n\right)\partial_n \tilde{\xi}_k\right|^{X_2}_{X_1} - \left.s^k \tilde{\xi}_k \right|^{X_2}_{X_1}\,.
\label{g22}
\end{align}
These transformation laws could be matched by a different type of charged particles whose transformations depend on a gauge vector $\xi_j$ rather than a scalar function. See \cite{Seiberg:2020wsg,Seiberg:2020cxy} for interpretations in term of ``lineons'', particles restricted to move in lines. 

We remark that all the known explicit models where these tensor theories are charged break some Lorentz invariance and some spacial symmetries. Then these ways of completing the graviton theory should, as expected, give up relativistic invariance.      

\section{Discussion}\label{DS}

We have found generalized symmetries for the graviton theory, associated with the failure of Haag duality for regions with non-trivial homotopy group $\pi_1$, containing non-contractible loops. This is the same setup of generalized symmetries for ordinary gauge models. However, the charges of the non-local operators for the graviton are not invariant under Lorentz transformations. This prompts the following discussion. A more detailed account will be published in a forthcoming paper. 

Global charges with Lorentz indices are forbidden by the Coleman-Mandula theorem, based on $S$ matrix properties. In the present framework for generalized symmetries, there is a very simple reason why a theory cannot have a mixing of Lorentz indices with the generalized symmetry labels, a sort of generalization of the Coleman-Mandula theorem. This is the existence of a stress tensor. In the presence of a stress tensor, we can form a local twist for Poincare symmetries that implements the symmetries on the region $R$ and not in the complement. On the other hand, a Poincare twist generated by the stress tensor is an additive operator in $R$. Therefore, if a non-local operator is Poincare transformed (infinitesimally) by the twist, both, the operator and its transformation, must belong to the same non-local class. This forbids Lorentz indices in the class labels. It also shows the limitations of models where this mixing of Lorentz and symmetry charges occur.  

It remains to show that the twist in $R$ implements the transformation on the non-local operator. But this follows from the fact that the Poincare twist in a ball $B$ that includes $R$ does implement the transformation on the non-local operator, which is an additive operator in the ball. This twist is the product of a twist in $R$ and one in the complement of $R$ inside $B$, and this later commutes with the non-local operator. Therefore, the twist in $R$ must transform the non-local operators.   

This argument indicates a reason allowing for the difference between ordinary gauge theories and the graviton field. In this latter theory, there is no stress tensor by the Weinberg-Witten theorem. Or, put in another way, we can derive the Weinberg-Witten theorem in this case by the existence of generalized symmetries mixed with Lorentz indices. As we have seen, this is incompatible with the existence of a stress tensor. Variations of the present ideas in non-relativistic models could be useful to show the necessity of breaking of rotational symmetry for certain fracton models.

\section*{Acknowledgements} 
The work of H. C.  is partially supported by CONICET, CNEA, and Universidad Nacional de Cuyo, Argentina, and an It From Qubit grant by the Simons Foundation.  The work of J.M is supported by a DOE QuantISED grantDE-SC0020360 and the Simons Foundation It From Qubit collaboration (385592). The work of V. B.  is supported by CONICET, Argentina.

\appendix

 \section{Linarized gravity in the ADM approach \label{ADM}}
Here we present the linearized gravity theory of section (\ref{LG}) using the usual ADM approach first presented in \cite{ADM}.  We follow the notation and general ideas in \cite{BlauNotes}. The quadratic ADM action that  is equivalent to the Einstein-Hilbert action and the Fierz-Pauli action up to boundary terms can be written as
\be
S_{ADM}= \int dt\,d^3x \, \left[\left(1+\frac{h}{2}- \frac{h_{00}}{2}\right) {}^3R^{(1)} + {}^3R^{(2)} +K_{(1)}^{ij}K^{(1)}_{ij}-K_{(1)}^2\right]\,,
\label{SADM}
\ee
where ${}^3R$ is the three-dimensional curvature, and the foliation is then defined by the lapse function $N$ and shift vector $\mathcal{N}^i $
\be 
N = \sqrt{(1-h_{00}) }+\mathcal{O}(h^2)=1-\frac{h_{00}}{2}+\mathcal{O}(h^2)\,, \quad \mathcal{N}^i = - h^{0i}+\mathcal{O}(h^2)\,,
\label{Foliation}
\ee
that corresponds to the extrinsic curvature defined as
\be 
K_{ij}=\frac{1}{2}\left( \dot{h}_{ij} - \partial_i h_{0j} - \partial_j h_{0i}   \right)\,.
\label{ExtrinsicCurvature}
\ee
The foliation determined by $N$ and ${\cal N}^i$ and the induced  metric $g_{ij}=\delta_{ij}+h_{ij}$ recovers $g_{\mu\nu}=\eta_{\mu\nu} + h_{\mu\nu}$ from the ADM metric 
\bea 
ds^2 &=& (-N^2 +  q_{ij}\, \mathcal{N}^i \mathcal{N}^j ) dt^2 +q_{ij}\,\mathcal{N}^j dx^i dt + q_{ij}\,\mathcal{N}^i dx^j  dt  + q_{ij} \,dx^i dx^j =\nonumber \\ &=& -(1-h_{00}) dt^2 + 2h_{0b} dt \,dx^b  +( \delta_{ab}+ h_{ab})dx^a dx^b  + \mathcal{O}(h^2)\,.
\eea
The standard ADM momenta and constraints reproduce the correct momenta  (\ref{Momenta}) and the constraints (\ref{Constraints}) presented in section (\ref{LG})
\bea
&{}& \pi_{ij} = \sqrt{q}(K_{ij}-q_{ij}K)= \frac{1}{2}\left( \dot{h}_{ij} - \partial_i h_{0j} - \partial_j h_{0i} - \delta_{ij} \dot{h}^{k}_{\,\,k} + 2\delta_{ij} \partial^k h_{0k}  \right) +\mathcal{O}(h^2)\,,\qquad \label{MomentaADM}\\
&{}&\mathcal{H} = \sqrt{q} \left( -{}^3R +K_{ab}K^{ab}-K^2\right) =  \partial_a \partial_a h_{bb}- \partial_a \partial_b h_{ab}  + \mathcal{O}(h^2)\,,\qquad\label{HConstraint}\\
&{}& \mathcal{H}_i = \sqrt{q} \,\overline{\nabla}^j \pi_{ij}= -2 \partial^j \pi_{ij} + \mathcal{O}(h^2)\,,\qquad \label{PConstraint}
\eea
where $\overline{\nabla}_i$ is the projected covariant derivative over the Cauchy slice of Minkowski space-time with foliation (\ref{Foliation}). This acts over $\mathcal{O}(h)$ variables as the usual derivative $\partial_i$.

In addition,  the ADM hamiltonian can be recovered by Legendre transforming (\ref{SADM}) 
\begin{align}
H_{ADM}=& \int d^3x\, \left[N(x)\mathcal{H}(x)+\mathcal{N}^i(x)\mathcal{H}_i(x)\right]
\nonumber \\
=& \int d^3x\, \left[\mathcal{H}^{(2)}+\left(1-\frac{h_{00}}{2}\right)\mathcal{H}^{(1)}(x)-h^{0i}(x)\mathcal{H}^{(1)}_i(x)\right]\,. \label{HamiltonianADM}
\end{align} 
In other words, the linearized gravity Hamiltonian without boundary terms is not a pure linear combination of constraints.  However,  the constraints coming from the lapse function and shift vector are equivalent to the ones derived in section (\ref{LG}) from the Lagrange multipliers $h_{00}$ and $h_{0i}$ respectively, as one would expect from the choice (\ref{Foliation}).

\noindent
It will be useful to keep in mind that we might compute the Heisenberg equations for the canonical variables using (\ref{HamiltonianADM})
\bea
i \dot{h}_{ij}(x) &= &\left[ h_{ij}(x), H_{ADM}\right] \quad \Rightarrow\quad \pi_{ij} =  \frac{1}{2}\left( \dot{h}_{ij} - \partial_i h_{0j} - \partial_j h_{0i} - \delta_{ij} \dot{h}^{k}_{\,\,k} + 2\delta_{ij} \partial^k h_{0k}  \right)\,, \nonumber\\
i \dot{\pi}_{ij}(x)&=& \left[ \pi_{ij}(x), H_{ADM}\right]  \quad \Rightarrow\quad G^{(1)}_{ij} =0\,.
\label{Heisenberg}
\eea

\section{Flux commutator for the case of two squares \label{Squares}}
\begin{figure}[h]
    \centering
    \includegraphics[width=0.65\textwidth]{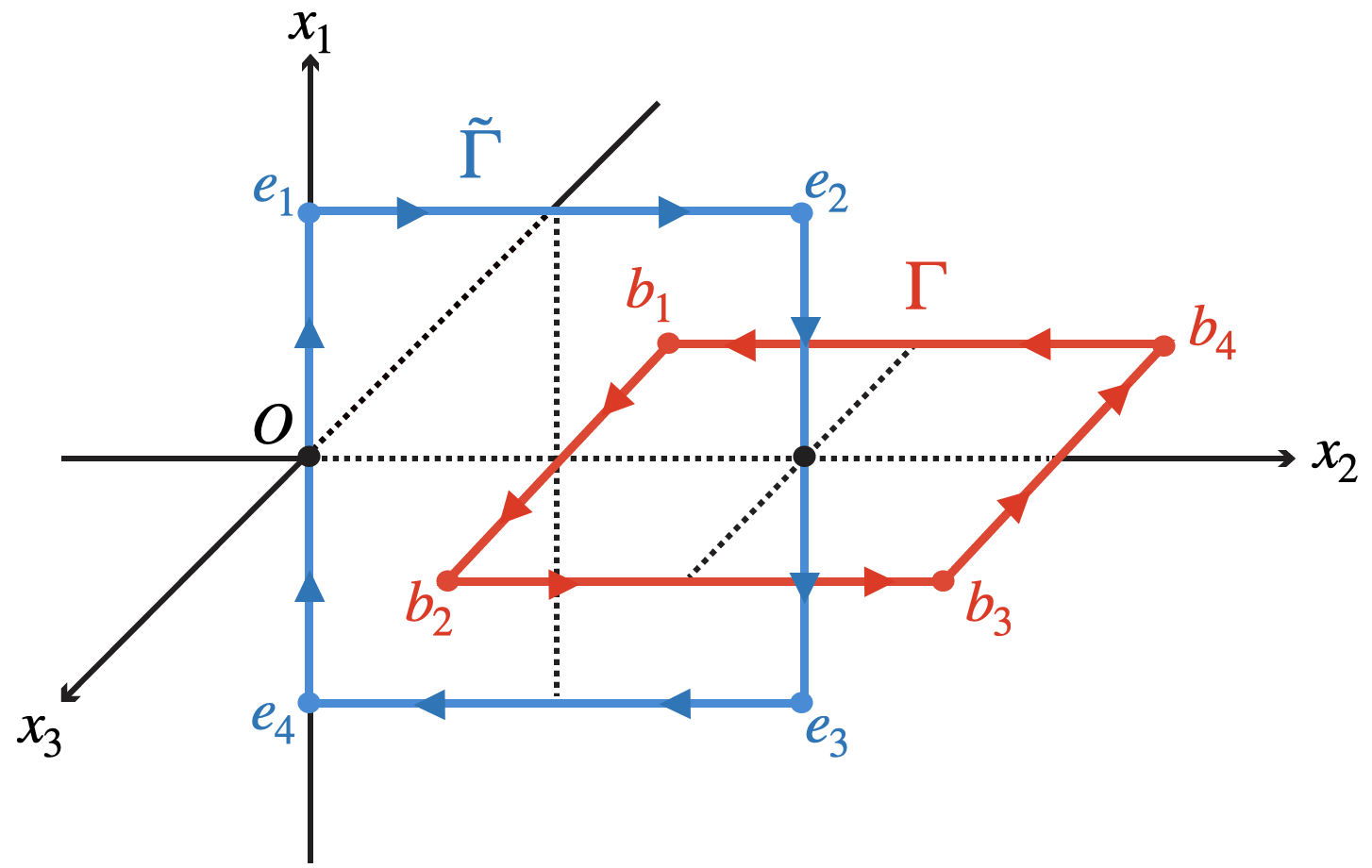}
    \caption{Regions $S$ and $\tilde{S}$ delimited by $\Gamma$ and  $\tilde{\Gamma}$ for $e=x_3=0$ and $b=y_1=0$.}
    \label{Flujos}
\end{figure}

For the Maxwell field in $(3+1)$  dimensions, we have that the commutation relations between the electric and magnetic fields  are given by  (\ref{MaxwellCommutator}). The commutator of the fluxes $\Phi^B$ and  $\Phi^E$ associated with the bi-dimensional surfaces $S$ and $\tilde{S}$  respectevely  can be computed as
\bea
\left[\Phi^B,\Phi^E\right] &=& \int_{S_E}\int_{S_B}\left[B_j(y),E_i(x)\right] dS^j(y)\,d\tilde{S}^i(x)=\nonumber\\
&=& i \int_{S_E}\int_{S_B} \varepsilon_{jik}\left[\partial^k \delta(x-y)\right] dS^j(y)\,d\tilde{S}^i(x)\,.
\label{CommutatorSquaresMaxwell} 
\eea
Lets suppose that the curve that defines the region $S$ is given by $\Gamma=\partial S = \{b_1,\,b_2,\,b_3,\,b_4 \}$ and that the curve  that defines $\tilde{S}$ is denoted by $\tilde{\Gamma}=\partial \tilde{S} = \{e_1,\,e_2,\,e_3,\,e_4 \}$, where 
\begin{align}
& b_1\left(0,\,\frac{L}{2},\,-\frac{L}{2}\right)\,,\quad  
b_2\left(0,\,\frac{L}{2},\,\frac{L}{2}\right)\,,\quad 
\,\,\,b_3\left(0,\,\frac{3L}{2},\,\frac{L}{2}\right)\,,\quad 
b_4\left(0,\,\frac{3L}{2},\,-\frac{L}{2}\right)\,,
\label{MagneticSquare} \\
& e_1\left(\frac{L}{2},\,0,\,0\right)\,,\quad \quad e_2\left(\frac{L}{2},\,L,\,0\right)\,, \quad \,\,\, e_3\left(-\frac{L}{2},\,L,\,0\right)\,,\quad e_4\left(-\frac{L}{2},\,0,\,0\right)\,. \label{ElectricSquare} 
\end{align}
This geometry is depicted in the figure \ref{Flujos}. The vectors that are normal to the surfaces  $S$ and  $\tilde{S}$ (associated with the circulation we have drawn) yield $n_i =\delta_{i1} $ and $ \tilde{n}_i=-\delta_{i3}$. Therefore surface differentials are given by
\bea
dS_j(y) &=& n_j\,dy_2 dy_3=\delta_{j1}dy_2 dy_3\,,\,\,\,\,\quad y_2 \in\,\,\, \left[ \frac{L}{2} ,  \frac{3L}{2}  \right]\,,\quad y_3 \in \,\left[- \frac{L}{2} ,  \frac{L}{2}   \right]\,,\\
\label{SurfaceDifferentialMagnetic}
d\tilde{S}_i(x) &=& \tilde{n}_i\,dx_1 dx_2=-\delta_{i3}dx_1 dx_2 \,,\quad x_1 \in \left.\left.\left[-\frac{L}{2}, \frac{L}{2} \right]\,,\quad x_2 \in \right[0, L \right]\,. \label{SurfaceDifferentialElectric} 
\eea
\noindent
Replacing equations (\ref{SurfaceDifferentialElectric}) and (\ref{SurfaceDifferentialMagnetic}) in (\ref{CommutatorSquaresMaxwell}),  we have that the commutator yields
\be
[\Phi^E,\Phi^B] = i \int^\frac{L}{2}_{-\frac{L}{2}}\int_0^L\int^\frac{3L}{2}_{\frac{L}{2}}\int^\frac{L}{2}_{-\frac{L}{2}}  \delta(x_1)\,\delta(-y_3)\,\partial_2 \, \delta(x_2-y_2) \, dx_1 \,dx_2 \,dy_2 \,dy_3 = i\,.
 \ee
 Furthermore, one can check that if the squares are rotated the result remains unchanged (except for a global minus sign related to the orientation of the surfaces). However, if one separates the squares and their areas no longer touch the result is always zero.  This is because the commutator $[\Phi^B,\Phi^E] $ is proportional to the linking number between the curves $\Gamma$ and $\tilde{\Gamma}$.

The same process can be done for the graviton considering the commutator  (\ref{ElectricMagneticCommutator}).  We want to compute the commutators of the fluxes defined by (\ref{MagneticFluxes}) and (\ref{ElectricFluxes})  in the set-up presented in figure \ref{Flujos}.  Specifically,  we may use the notation
\bea 
\Phi^B_G = \int_{S} B_{ij}(x) \, g^j (x) \,dS^i\,,\quad g_j (x) = a_j\,,\,\, -s_{ j n}x_n \,,\,\, k x_j \,,\,\,(b_j x_nx_n -2x_j b_n x_n)\,,\\
\Phi^E_F = \int_{\tilde{S}}\, E_{ij}(x) f^j(x) \, d\tilde{S}^i\,,\quad  f_j (x)= \tilde{a }_j\,,\,\,- \tilde{s}_{ j n}x_n \,,\,\, \tilde{k} x_j \,,\,\,(\tilde{b}_j x_nx_n -2x_j \tilde{b}_n x_n)\,. 
\eea
For the  inter-located square geometries defined  above we may write
\bea
\left[\Phi^B_G , \Phi^E_F \right] &=& \int_{S} \int_{\tilde{S}} \left[ E_{kl}(y), B_{ij}(x)\right] \,g^l(y)\,f^j(x)\, dS^k (y) \,d\tilde{S}^i (x)\label{CommutatorSquaresGraviton} \\
&=& \int^\frac{L}{2}_{-\frac{L}{2}}\int_0^L\int^\frac{3L}{2}_{\frac{L}{2}}\int^\frac{L}{2}_{-\frac{L}{2}} \left[E_{3i}(x) , B_{1j}(y) \right]\, f^i(x)g^j(y)\,\, dx_1 \,dx_2 \,dy_2\, dy_3 \,.
\nonumber
\eea
From here,  computing case by case, we get the expected results 
\bea
&{}&   \left[ \Phi^B_P  ,\Phi^E_K \right]  = i\,  a^i\,\tilde{b}_i  \,, \quad   \left[ \Phi^B_J  ,\Phi^E_J \right]=\frac{i}{4} \,s^{ij}\,\tilde{s}_{ji} \,,\nonumber \\ &{}&  \left[ \Phi^B_D  ,\Phi^E_D \right]=\frac{i}{2}\, \kappa\,\tilde{\kappa} \,, \quad \left[ \Phi^B_K  ,\Phi^E_P \right]=i\,b^i\, \tilde{a}_i  \,.
\eea
We note that this is still true if we move the squares around changing the coordinates $x_3$ and $y_1$ in the range $[-L/2,L/2]$ (without changing the linking number) but the result vanishes if we separate the squares, which is a good symptom as we expect the general result to be topological.

\bibliographystyle{utphys}
\bibliography{EE}

\end{document}